\documentclass[11pt,a4paper]{article}

\usepackage{geometry}
\usepackage{ulem}
\usepackage{microtype}           
\usepackage{amsmath,amssymb,amsthm}
\usepackage{mathtools}
\usepackage{physics}
\usepackage{bm}
\usepackage{graphicx}
\usepackage{subfigure}
\usepackage{caption}
\usepackage{mathrsfs}
\usepackage{amsthm}
\usepackage[colorlinks=true,linkcolor=red,citecolor=blue,urlcolor=blue,bookmarks]{hyperref}
\usepackage{xcolor}

\newcommand{\affiliation}[1]{\\ \parbox[t]{0.9\textwidth}{\centering\small\textit{#1}}} 

\theoremstyle{plain}

\theoremstyle{definition}

\theoremstyle{remark}

\def \be {\begin{equation}}
\def \ee {\end{equation}}
\def \ba {\begin{array}}
\def \ea {\end{array}}
\def \bea{\begin{eqnarray}}
\def \eea{\end{eqnarray}}
\def \nn {\nonumber}

\def \mOz {\mathcal{O}_0}
\def \mOo {\mathcal{O}_1}
\def \mO {\mathcal{O}}
\title{Spacetime Density Matrix: Formalism and Properties}
\author{Wu-zhong Guo\footnote{wuzhong@hust.edu.cn}\\ \affiliation{School of Physics, Huazhong University of Science and Technology, \\
Luoyu Road 1037, Wuhan, Hubei 430074, China}}
\date{}

\begin{document}

\maketitle
\begin{abstract}
In this paper, we develop the general formalism and properties of the spacetime density matrix, which captures correlations among different Cauchy surfaces and can be regarded as a natural generalization of the standard density matrix defined on a single Cauchy surface. We present the construction of the spacetime density matrix in general quantum systems and its representation via the Schwinger–Keldysh path integral. We further introduce a super-operator framework, within which the spacetime density matrix appears as a special case, and discuss possible generalizations from this perspective. We also show that the spacetime density matrix satisfies a Liouville–von Neumann type equation of motion.
When considering subsystems, a reduced spacetime density matrix can be defined by tracing over complementary degrees of freedom. We study the general properties of its moments and, in particular, derive universal short-time behavior of the second moment. We find that coupling between subsystems plays a crucial role in obtaining nontrivial results. Assuming weak coupling, we develop a perturbative method to compute the moments systematically.

\end{abstract}

\tableofcontents

\section{Introduction}
The fundamental quantities in a quantum system are the correlation functions of operators, which can be directly related to experimental observations. In the framework of algebraic quantum field theory (QFT), the Wightman reconstruction theorem states that the collection of all Wightman functions is equivalent to the full specification of the theory itself, namely the reconstruction of the Hilbert space together with the global vacuum state \cite{Streater:1989vi}.

This theorem can be slightly generalized to subregions on a given Cauchy surface. If one knows all correlation functions of operators restricted to a subregion, it should in principle be possible to reconstruct the reduced density matrix of that subregion. Hence, the reduced density matrix captures the correlation information of spacelike-separated subregions on a Cauchy surface. From the reduced density matrix one may define entropy-related quantities, among which the entanglement entropy (EE) plays a central role. EE has found broad applications in quantum many-body physics, quantum field theory, and gravity; see the review articles and references therein \cite{Amico:2007ag,Calabrese:2009qy,Takayanagi:2025ula}.

In the usual approach, one chooses a time slice $C_0 : t = t_0$ and defines the density matrix $\rho_0$. The expectation value of operators $\mathcal{O}_0$ localized on $C_0$ is then given by $\Tr(\rho_0 \mathcal{O}_0)$. One can also consider the expectation value of an operator located on a later time slice $C_1 : t = t_1$, defined as $\mathcal{O}_1(t_1)=U(t_1,t_0)^\dagger \mathcal{O}_1 U(t_1,t_0)$, where $U$ is the time-evolution operator and $\mathcal{O}_1$ is the corresponding operator at $C_0$. Its expectation value reads
\bea\label{one_point_time_t}
\Tr(\rho_0  \mathcal{O}_1(t_1)) = tr(\rho(t_1) \mathcal{O}_1),
\eea
where $\rho(t_1) := U(t_1,t_0) \rho_0  U(t_1,t_0)^\dagger$. The right-hand side makes explicit use of the Schrödinger picture for the evaluation of expectation values.

If one is interested in the correlation functions between operators $\mathcal{O}_0$ and $\mathcal{O}_1(t_1)$ located on different timeslices, one should compute $\Tr(\rho_0 \mathcal{O}_0 \mathcal{O}_1(t_1))$. Similar to relation (\ref{one_point_time_t}), it is possible to introduce an operator $T_{C_0C_1}$ such that
\bea\label{Spacetime_operator}
\Tr(\rho_0 \mathcal{O}_0 \mathcal{O}_1(t_1)) = tr\left(T_{C_0C_1}\mathcal{O}_0 \mathcal{O}_1\right).
\eea
This idea was recently proposed in \cite{Milekhin:2025ycm}, motivated in part by the attempt to define the notion of \textit{entanglement in time}. A similar concept also appeared in \cite{Cotler:2017anu}, where the authors introduced superdensity operators to study quantum information in spacetime. The operator $T_{C_0C_1}$ can be reconstructed in any quantum system, as we will demonstrate below. By definition, it encodes the correlation information between different Cauchy surfaces $C_0$ and $C_1$.

Notice that by taking $\mathcal{O}_0 = I$, Eq.(\ref{Spacetime_operator}) reduces to Eq.(\ref{one_point_time_t}). Thus, $T_{C_0C_1}$ can be regarded as a generalization of the density matrix from a single Cauchy surface to spacetime regions. In this sense, $T_{C_0C_1}$ plays a role analogous to the density matrix, and we may refer to it as a spacetime density matrix. For subsystems, one may also define a reduced form of $T_{C_0C_1}$, which allows for the construction of entropy-related quantities. Such quantities are crucial for understanding the notion of entanglement between causally connected subsystems \cite{Olson:2010jy, Doi:2022iyj}.

In previous studies, entanglement in QFTs for spacelike subsystems has already provided deep insights into the entanglement structure of quantum field states and into holographic properties of gravity \cite{Ryu:2006bv}–\cite{Almheiri:2019qdq}. It is therefore natural to expect that extending the concept of entanglement to the temporal domain will shed new light on the dynamical and causal aspects of QFTs and holography in real time.

The motivation of this paper is to investigate the general formalism and properties of the spacetime density matrix. For a general quantum system with a given initial state and Hamiltonian, it is possible to construct the spacetime density matrix associated with arbitrary Cauchy surfaces. This operator can be represented using the Schwinger–Keldysh path integral \cite{Schwinger:1960qe,Feynman:1963fq,Keldysh:1964ud}, which is particularly useful for studying open quantum systems. The Schwinger–Keldysh formalism was employed in \cite{Dong:2016hjy} in the context of proving the Hubeny–Rangamani–Takayanagi (HRT) formula \cite{Hubeny:2007xt}. In this sense, the spacetime density matrix and its reduced form can be viewed as a natural extension of the Schwinger–Keldysh representation of time-dependent density matrices.

We find it useful to introduce a super-operator that maps operators to operators. The spacetime density matrix can be regarded as a special case of such a mapping. With this formalism, one can further generalize to other operators. In this work, we define a dual operator associated with $T_{C_0C_1}$ and introduce a generalization obtained by modifying the final state. We also derive the Liouville–von Neumann equation for the spacetime density matrix, which admits a more compact representation when expressed in terms of the super-operator.

The moments of the (reduced) spacetime density matrix play a key role in exploring its properties. In certain cases, the path-integral representation provides a practical tool to compute these moments. We further analyze their universal behavior in the short-time or weak-coupling limits, where they exhibit robust general features. We propose that the spacetime density matrix serves as a useful diagnostic of the dynamical evolution of a system and its interaction with other systems. To this end, we develop a perturbative method for evaluating the moments of (reduced) spacetime density matrices and illustrate its application through explicit examples.

\section{Spacetime density matrix and path integral representation}
Consider a quantum system with Hamiltonian $H$. The time-evolution operator is defined as
$U := e^{-i H t}$. We consider two different Cauchy surfaces at time slices $t=t_0$ and $t=t_1$, associated with the Hilbert spaces $\mathcal{H}_0$ and $\mathcal{H}_1$, respectively. These Hilbert spaces are isometric to each other.

The initial density matrix of the system at $t_0$ is denoted by $\rho_0$. For operators $\mOz, \mOo$ defined on the Cauchy surface at $t=t_0$, the corresponding correlation function is
$ \Tr(\rho_0 \mOz \mOo) $. Under time evolution, the operator $\mOo$ can be mapped to the Cauchy surface at $t=t_1$ as
$\mOo(t_1) := U(t_1,t_0)^\dagger \mOo U(t_1,t_0)$. Accordingly, the anti-time-ordered correlation function can be defined as
\bea\label{Timelike_correlator}
\langle \mOz \mOo(t_1) \rangle_{\rho_0} := \Tr\left[\rho_0 \mOz \mOo(t_1)\right].
\eea
Now we would like to show that there exists a unique operator
\bea
T_{C_0C_1}:\mathcal{H}_0\otimes \mathcal{H}_1\to \mathcal{H}_0\otimes \mathcal{H}_1,
\eea
which satisfies\footnote{In the following, we use $\Tr$ to denote the standard trace on a single Hilbert space, while $tr$ denotes the trace on the tensor product Hilbert space $\mathcal{H}_0\otimes \mathcal{H}_1$, or more generally on $\mathcal{H}_0\otimes \mathcal{H}_1\otimes \cdots \otimes \mathcal{H}_{N-1}$.}
\bea\label{OperatorT_definition}
\langle \mOz \mOo(t_1) \rangle{\rho_0} = tr \left( T_{C_0 C_1} \mOz \mOo \right).
\eea
Here the trace on the right-hand side is taken over the Hilbert spaces $\mathcal{H}_0$ and $\mathcal{H}_1$. We choose a basis ${ |i\rangle }$ for the Hilbert space\footnote{In the derivation we assume the Hilbert space is discrete; however, the result can be straightforwardly generalized to the continuous case.}. The operators $T_{C_0C_1}$, $\mOz$, and $\mOo$ can then be expanded as follows\footnote{For simplicity, we omit the explicit $\sum$ notation. Repeated indices are implicitly summed over, following the Einstein summation convention familiar from tensor analysis.}:
\bea
&&T_{C_0C_1}=t_{mn,st}|m\rangle \langle n| \otimes |s\rangle \langle t|, \nn \\
&&\mOz= a_{0,ij}|i\rangle \langle j|,\quad \mOo=a_{1,kl}|k\rangle \langle l|.
\eea
The right-hand side of (\ref{OperatorT_definition}) can be written as
\bea\label{T_form}
&&tr (T_{C_0 C_1} \mOz \mOo)
= t_{mn,st}\langle n|\mOz|m\rangle \langle t|\mOo |s\rangle \nn \\
&&\phantom{tr (T_{C_0 C_1} \mOz \mOo)}
= t_{ji,lk}, a_{0,ij}, a_{1,kl}.
\eea
On the other hand, the right-hand side of (\ref{Timelike_correlator}) is given by
\bea\label{Timecorrelator_form}
&&\langle \mOz \mOo(t_1) \rangle_{\rho_0}
= a_{0,ij} a_{1,kl}tr(\rho_0 |i\rangle \langle j|U^\dagger|k\rangle \langle l|U ) \nn \\
&&\phantom{\langle \mOz \mOo(t_1) \rangle_{\rho_0}}
= a_{0,ij} a_{1,kl} \langle l|U \rho_0 |i\rangle \langle j|U^\dagger|k\rangle.
\eea
By comparing (\ref{T_form}) with (\ref{Timecorrelator_form}), we obtain $t_{ji,lk}$. Thus, the operator $T_{C_0C_1}$ is given by
\bea\label{Transition_two}
T_{C_0C_1} = \langle l|U \rho_0 |i\rangle \langle j|U^\dagger|k\rangle  |j\rangle \langle i| \otimes |l\rangle \langle k|,
\eea
where $U := U(t_1,t_0)$. Using the completeness relation of the basis $\sum |i\rangle \langle i| = I$, one can also write
\bea\label{Two_Form1}
T_{C_0C_1} = U^\dagger |k\rangle \langle l| U \rho_0 \otimes |l\rangle \langle k|.
\eea
Alternatively, it can be expressed in the equivalent form
\bea\label{Two_Form2}
T_{C_0C_1} = |j\rangle \langle i| \otimes U \rho_0 |i\rangle \langle j| U^\dagger.
\eea
In the derivation of the expression for $T_{C_0C_1}$ (\ref{Transition_two}), we have implicitly assumed that the basis on the Cauchy surface at $t=t_1$ is the same as that at $t=t_0$. This assumption is not necessary; however, the two bases must be unitarily equivalent. Any other choice of basis on the Cauchy surface at $t=t_1$ would yield an operator that is related to (\ref{Transition_two}) by a unitary transformation.


As we have shown above, the definition of $T_{C_0C_1}$ involves the information of correlation functions for operators located on different Cauchy surfaces, $t=t_0$ and $t=t_1$. It therefore contains more information than the density matrix $\rho_0$ of the system. In the following, we will show that it reduces to the density matrix $\rho$ in certain specific cases. For this reason, it is appropriate to refer to this operator as a \textit{spacetime density matrix}, since it extends beyond a single Cauchy surface. Alternatively, one may refer to it as a \textit{transition operator} between the Hilbert spaces $\mathcal{H}_0$ and $\mathcal{H}_1$.

Let us now discuss some basic properties of $T_{C_0C_1}$.
It can be verified that $T_{C_0C_1}$ indeed satisfies the condition (\ref{OperatorT_definition}). From the form (\ref{Two_Form2}), we also have
\bea
I \otimes U(t_2,t_1) T_{C_0C_1} I \otimes U(t_2,t_1)^\dagger = T_{C_0C_2},
\eea
where $C_2$ denotes the Cauchy surface at $t=t_2$. Therefore, any spacetime density matrices corresponding to two Cauchy surfaces are unitarily equivalent.
Another simple property is
\bea\label{T12_partial_trace}
tr_{C_1} T_{C_0C_1} = \rho_0, \quad
tr_{C_0} T_{C_0C_1} = \rho(t_1) := U(t_1,t_0), \rho_0, U(t_1,t_0)^\dagger.
\eea
This shows that the density matrices $\rho_0$ at $t=t_0$ and $\rho(t_1)$ at $t=t_1$ are special cases of the spacetime density matrix $T_{C_0C_1}$. This can also be seen directly from the definition of $T_{C_0C_1}$: if we take the operator $\mOo$ or $\mOz$ to be the identity $I$, the definition of $T_{C_0C_1}$ reduces to the definition of $\rho_0$ and $\rho(t_1)$, respectively.

By the same motivation, we can also consider the time-ordered correlator
\bea
\langle \mOo(t_1)\mOz\rangle_{\rho_0}:=\tr(\rho_0 \mOo(t_1) \mOz).
\eea
We introduce the spacetime density matrix $T'_{C_0C_1}$ satisfying
\bea\label{THJ_definition}
\langle \mOo(t_1)\mOz\rangle{\rho_0} = \tr \big( T'_{C_0C_1} \mOz \mOo \big).
\eea
By similar method, we can prove
\bea\label{T01_HJ_expression}
T'_{C_0C_1} = \langle k |U|j\rangle \langle i|\rho_0 U^\dagger |l\rangle |i\rangle\langle j|\otimes |k\rangle\langle l|.
\eea
It can also be written as
\bea
T'_{C_0C_1} = \rho_0 U^\dagger |l\rangle \langle k|U \otimes |k\rangle \langle l|,
\eea
or equivalently,
\bea
T'_{C_0C_1} = |i\rangle \langle j| \otimes U|j\rangle \langle i|\rho_0 U^\dagger.
\eea
By comparing with (\ref{Two_Form1}) and (\ref{Two_Form2}), it is clear that
\bea
T'_{C_0C_1} = T_{C_0C_1}^\dagger.
\eea
This result follows from the fact that $\langle \mOz \mOo(t_1)\rangle_{\rho_0}^* = \langle \mOo(t_1)\mOz\rangle_{\rho_0}$ for Hermitian operators $\mOz$ and $\mOo$. Moreover, $T_{C_0C_1}^\dagger$ also satisfies
\bea\label{property_T01_HJ}
\tr_{C_1} T_{C_0C_1}^\dagger = \rho_0, \quad \tr_{C_0} T_{C_0C_1}^\dagger = U \rho_0 U^\dagger = \rho(t_1).
\eea

\subsection{Representation by Schwinger-Keldysh path integral}\label{section_SK}

\begin{figure}[htbp]
    \centering
    \subfigure{
        \includegraphics[width=0.25\textwidth]{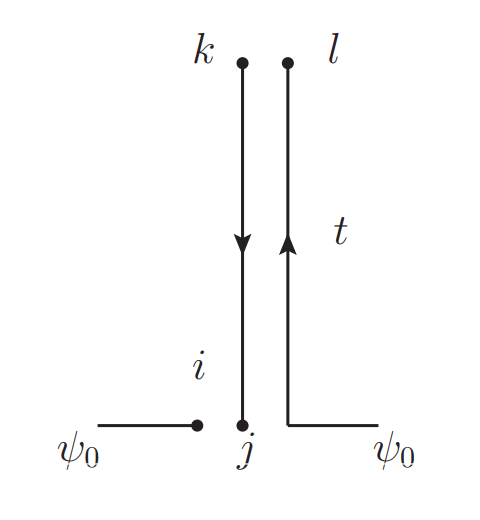}}
    \subfigure{
        \includegraphics[width=0.23\textwidth]{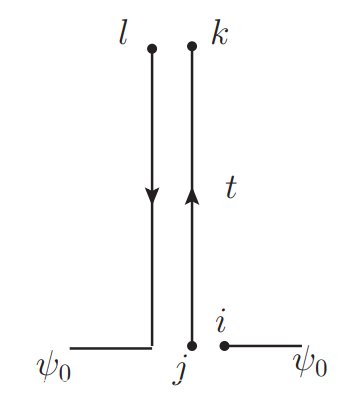}}
        \caption{Schwinger–Keldysh representation of the spacetime density matrix $T_{C_0C_1}$ (left) and its Hermitian conjugate $T_{C_0C_1}^\dagger$ (right). The initial state is denoted by $\psi_0$, while $i,j,k,l$ specify the boundary conditions. The arrows indicate the direction of time evolution.}\label{Fig_SK_T01}
\end{figure}

The spacetime density matrix can be represented by a Schwinger–Keldysh path integral with cuts. In particular, the expression (\ref{Transition_two}) of $T_{C_0C_1}$ can be translated into a path-integral form, see the review \cite{Haehl:2024pqu} for the introduction of Schwinger-Keldysh path integral. For simplicity, we assume that the density matrix is pure, $\rho_0=|\psi_0\rangle \langle \psi_0|$. The generalization to a mixed state is straightforward, and the thermal state will be discussed later. In this case we obtain
\bea\label{Transition_two_pure}
T_{C_0C_1} = \langle l|U|\psi_0\rangle  \langle \psi_0|i\rangle \langle j|U^\dagger|k\rangle  |j\rangle \langle i|\otimes |l\rangle \langle k| .
\eea
Each term in the above expression admits a path-integral representation. The initial state $|\psi_0\rangle$ can be prepared by a Euclidean path integral. The factor $\langle l|U|\psi_0\rangle$ corresponds to real-time evolution from the initial state $|\psi_0\rangle$ to the basis state $|l\rangle$, while $\langle j|U^\dagger|k\rangle$ corresponds to backward time evolution from $|k\rangle$ to $|j\rangle$. Therefore, the operator $T_{C_0C_1}$ can be represented diagrammatically as shown in Fig.~\ref{Fig_SK_T01}, which depicts a typical Schwinger–Keldysh contour with cuts at $t=t_0$ and $t=t_1$, with boundary conditions labeled by $i,j$ and $k,l$, respectively. One may also refer to \cite{Dong:2016hjy} \cite{Colin-Ellerin:2020mva,Colin-Ellerin:2021jev} for the construction of the path-integral representation of $\rho(t)$, which is helpful for understanding the case of $T_{C_0C_1}$ here.

\begin{figure}[htbp]
    \centering
    \subfigure{
        \includegraphics[width=0.48\textwidth]{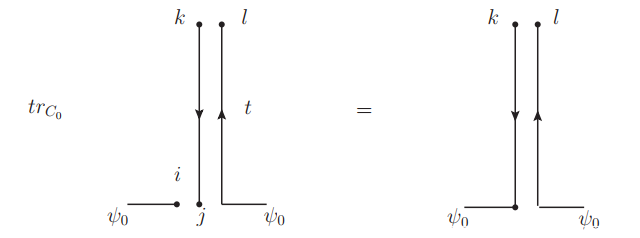}}
    \subfigure{
        \includegraphics[width=0.48\textwidth]{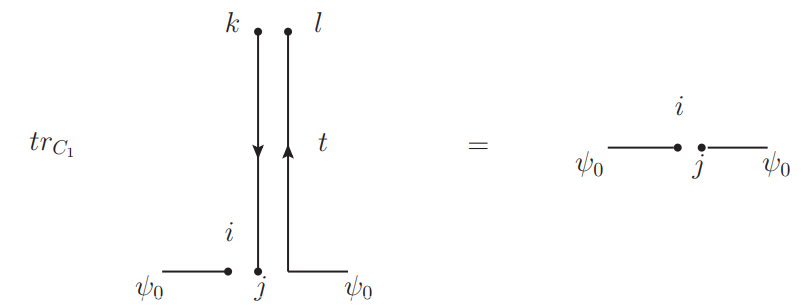}}
        \caption{Illustration of the properties (\ref{T12_partial_trace}) of the transition operator $T_{C_0C_1}$. The left panel shows the partial trace over $C_0$, which identifies the boundary conditions $i$ and $j$. The right panel shows the partial trace over $C_1$, corresponding to the identification of $k$ with $l$, and the forward and backward contributions cancel each other.}\label{Fig_SK_property}
\end{figure}

We can examine the properties (\ref{T12_partial_trace}) of $T_{C_0C_1}$ using its path integral representation. Taking a partial trace over $C_0$ or $C_1$ corresponds to identifying the boundary indices $i$ with $j$ or $k$ with $l$, respectively. This immediately leads to the relations in (\ref{T12_partial_trace}), as illustrated in Fig.~\ref{Fig_SK_property}. Performing a full trace over both $C_0$ and $C_1$ reduces the construction to the standard Schwinger–Keldysh path integral. It is also evident that inserting the operators $\mOz$ and $\mOo$ into the path integral at $C_0$ and $C_1$, respectively, yields the anti-time-ordered correlation function $\mathrm{tr}(\rho_0 \mOz \mOo(t_1))$, which is consistent with the definition of $T_{C_0C_1}$ in (\ref{OperatorT_definition}).

We can also represent the operator $T_{C_0C_1}^\dagger$. Assuming again a pure initial density matrix $\rho_0 = |\psi_0\rangle\langle\psi_0|$, the path-integral representation of $T_{C_0C_1}^\dagger$, derived from the expression (\ref{T01_HJ_expression}), is illustrated in Fig.~\ref{Fig_SK_T01}.
We can also show $T_{C_0C_1}^\dagger$ satisfies the properties (\ref{THJ_definition}) and (\ref{property_T01_HJ}).

\begin{figure}[htbp]
    \centering
    \subfigure{
        \includegraphics[width=0.23\textwidth]{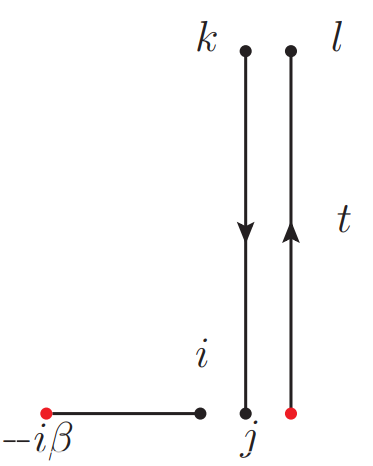}}
    \subfigure{
        \includegraphics[width=0.22\textwidth]{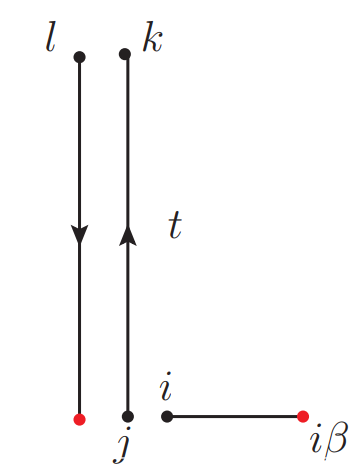}}
        \caption{Figure illustrating $T_{C_0C_1}$ (left) and $T_{C_0C_1}^\dagger$ (right) for the thermal initial state $\rho_0 = e^{-\beta H}$. The red dots indicate the identified points. }\label{Fig_SK_T01_thermal}
\end{figure}

For a thermal state $\rho_0 = e^{-\beta H}$, the Schwinger–Keldysh path integral with cuts can also be employed, as illustrated in Fig.~\ref{Fig_SK_T01_thermal}.

\subsection{Generalization to mutiple timeslices}\label{section_GMT}
The above construction can be generalized to multiple Cauchy surfaces.
For a Cauchy surface at $t=t_i$ with $i=0,1,\dots,N-1$, we associate a Hilbert space $\mathcal{H}i$.
We are also interested in the general $N$-point anti-time-ordered correlation function,
\bea
\langle \mO_0 \mO_1(t_1) \cdots \mO_i(t_i) \cdots \mO_{N-1}(t_{N-1}) \rangle_{\rho_0}
:= \tr\Big(\rho_0 \mO_0 \mO_1(t_1) \cdots \mO_i(t_i) \cdots \mO_{N-1}(t_{N-1})\Big),
\eea
where $\mO_i(t_i) := U(t_i,t_0)^\dagger \mO_i U(t_i,t_0)$ denotes the operator on the Cauchy surface $t=t_i$.
Here we assume the time ordering $t_0 < t_1 < \cdots < t_{N-1}$. We now introduce a spacetime density matrix:
\bea
T_{C_0C_1\cdots C_i\cdots C_{N-1}}:\mathcal{H}_0\otimes \mathcal{H}_1\cdots \otimes\mathcal{H}_i\cdots \otimes \mathcal{H}_{N-1} \to \mathcal{H}_0\otimes \mathcal{H}_1\cdots \otimes\mathcal{H}_i\cdots \otimes \mathcal{H}_{N-1},
\eea
which satisfies the defining relation
\bea\label{definition_T0N}
tr (T_{C_0C_1\cdots C_i\cdots C_{N-1}}\mOz \mOo\cdots \mO_i\cdots \mO_{N-1})=\langle \mOz \mOo(t_1)\cdots\mO_N(t_i)\cdots\mO_N(t_{N-1})\rangle_{\rho_0}.
\eea
We can also straightforwardly obtain the explicit form of the transition operator, in analogy with the two-Cauchy-surface case (\ref{Transition_two}). The derivation follows the same steps as in the two-surface case. The result is
\bea\label{mutiple_transition}
&&T_{C_0C_1\cdots C_i\cdots C_{N-1}}=\langle l_{N-1} |U(t_{N-1},t_0)\rho_0 |k_{0}\rangle \langle l_{0}|U(t_{1},t_{0})^\dagger|k_1\rangle\langle l_{1}|\cdots\\
&&\phantom{T_{C_0C_1\cdots C_i\cdots C_{N-1}}=}\langle l_i|U(t_{i+1},t_i)^\dagger|k_{i+1}\rangle\cdots |\langle l_{N-2}|U(t_{N-1},t_{N-2})^\dagger|k_{N-1}\rangle \nn \\
&&\phantom{T_{C_0C_1\cdots C_i\cdots C_{N-1}}=}|l_0\rangle \langle k_0| \otimes |l_1\rangle \langle k_1|\otimes \cdots \otimes |l_i\rangle \langle k_i|\cdots\otimes |l_{N-1}\rangle \langle k_{N-1}|.
\eea
It is also possible to consider the time-ordered correlation function:
\bea
\langle \mO_{N-1}(t_{N-1})\cdots\mO_i(t_i)\cdots\mOo(t_1)\mOz\rangle_{\rho_0}:=\Tr(\rho_0 \mO_{N-1}(t_{N-1})\cdots\mO_i(t_i)\cdots\mOo(t_1)\mOz).
\eea
\begin{figure}[htbp]
    \centering
    \subfigure{
        \includegraphics[width=0.32\textwidth]{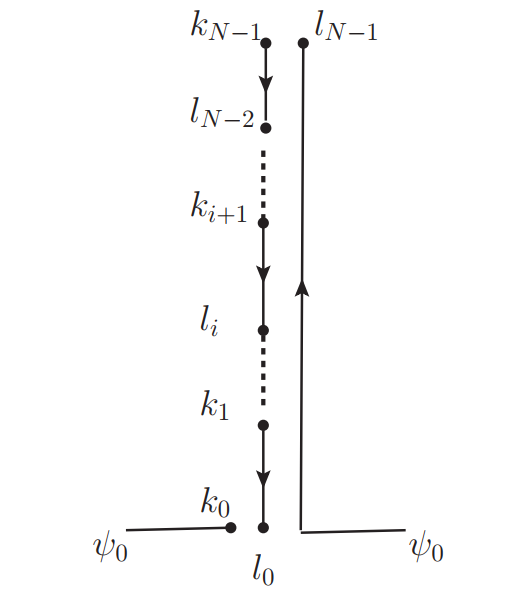}}
    \subfigure{
        \includegraphics[width=0.3\textwidth]{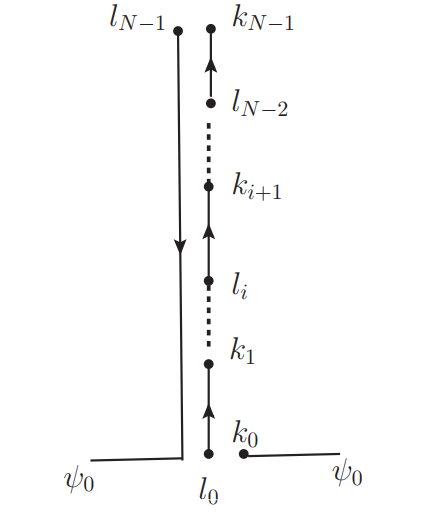}}
        \caption{Schwinger-Keldysh representation of the spacetime density matrix $T_{C_0C_1\cdots C_i\cdots C_{N-1}}$ (left) and its Hermitian conjugation $T_{C_0C_1\cdots C_i\cdots C_{N-1}}^\dagger$ (right).  }\label{Fig_SK_T0N}
\end{figure}

It can be shown that there exists an operator $T'_{C_0C_1\cdots C_i\cdots C_{N-1}}$ which satisifes 
\bea\label{definition_T0N_HJ}
tr (T'_{C_0C_1\cdots C_i\cdots C_{N-1}}\mOz \mOo\cdots \mO_i\cdots \mO_{N-1})=\langle \mO_{N-1}(t_{N-1})\cdots\mO_i(t_i)\cdots\mOo(t_1)\mOz\rangle_{\rho_0}.
\eea
Similar to the case of $T_{C_0C_1}$, we can show
\bea
T'_{C_0C_1\cdots C_i\cdots C_{N-1}}=T_{C_0C_1\cdots C_i\cdots C_{N-1}}^\dagger.
\eea
By using the expreesion (\ref{mutiple_transition}) for $T_{C_0C_1\cdots C_i\cdots C_{N-1}}$, we have
\bea
tr_{C_i}T_{C_0C_1\cdots C_i\cdots C_{N-1}}= T_{C_0C_1\cdots C_{i-1}C_{i+1}\cdots C_{N-1}}.
\eea
More generally, let us define $C_I=C_{i_1}C_{i_2}\cdots C_{i_I}$ with $i_k\in \{0,\cdots,N-1\}$ ($k=1,\cdots,I$),
\bea\label{partial_trace_T0N}
tr_{C_I}T_{C_0C_1\cdots C_i\cdots C_{N-1}}= T_{C_0C_1\cdots C_{i}\cdots C_{N-1}-C_I},
\eea
where $C_0C_1\cdots C_{i}\cdots C_{N-1}-C_I$ denotes the set obtained by removing $C_I$ from $C_0C_1\cdots C_{i}\cdots C_{N-1}$. As a special case, when $C_I=C_1\cdots C_{N-1}$ we have
\bea
tr_{C_1\cdots C_{N-1}}T_{C_0C_1\cdots C_i\cdots C_{N-1}}=\rho_0.
\eea
Similarly, when $C_I=C_0\cdots C_{N-2}$ we have
\bea
tr_{C_0\cdots C_{N-2}}T_{C_0C_1\cdots C_i\cdots C_{N-1}}=\rho_0(t_{N-1}).
\eea
It is more convenient to represent the transition operator $T_{C_0 C_1 \cdots C_i \cdots C_{N-1}}$ via the Schwinger–Keldysh path integral and its Hermitian conjugate, as illustrated in Fig.~\ref{Fig_SK_T0N}. One can also verify that it satisfies the properties (\ref{definition_T0N}), (\ref{definition_T0N_HJ}), and (\ref{partial_trace_T0N}).

\subsection{Reduced spacetime density matrix and entropy-related quantities}\label{Section_reduced_transition}

We have shown that the spacetime density matrix can be systematically constructed and expressed by using the path integral representation. In this construction, we consider the full Hilbert space at different timeslices or Cauchy surfaces. The transition operator $T_{C_0 C_1 \cdots C_i \cdots C_{N-1}}$ thus contains the information of correlation functions across the different time slices $C_0, \dots, C_{N-1}$.

For a density matrix $\rho$ and a given subsystem $A$, it is natural to consider the reduced density matrix $\rho_A := \tr_{\bar A} \rho$, where $\bar A$ denotes the complement of $A$. This idea can be extended to the spacetime density matrix. Assume that the Hilbert spaces $\mathcal{H}_i$ for $i=0,1,\dots,N-1$ can be factorized as
\bea
\mathcal{H}_i = \mathcal{H}_{i,A_i} \otimes \mathcal{H}_{i,\bar A_i},
\eea
where $A_i$ and $\bar A_i$ denote the subsystems on the $i$-th time slice. Then, the reduced transition operator for the combined subsystem $A := A_0 \cup A_1 \cup \cdots \cup A_{N-1}$ is defined as
\bea
T_{A_0 A_1 \cdots A_i \cdots A_{N-1}} := tr_{\bar A_0 \bar A_1 \cdots \bar A_i \cdots \bar A_{N-1}} T_{C_0 C_1 \cdots C_i \cdots C_{N-1}}.
\eea
As a result $T_{A_0A_1\cdots A_i \cdots A_{N-1}}$ can be seen as an operator
\bea
T_{A_0A_1...A_i...A_{N-1}}:\mathcal{H}_{A_0}\otimes \mathcal{H}_{A_1}\cdots\otimes\mathcal{H}_{A_i}\cdots\otimes \mathcal{H}_{A_{N-1}} \to \mathcal{H}_{A_0}\otimes \mathcal{H}_{A_1}\cdots\otimes\mathcal{H}_{A_i}\cdots\otimes \mathcal{H}_{A_{N-1}}.
\eea
For operators $\mathcal{O}_{A_i}$ acting on the Hilbert space $\mathcal{H}_{A_i}$, we also have
\bea
&&tr(T_{A_0A_1...A_i...A_{N-1}}\mO_{A_0}\mO_{A_1}\cdots\mO_{A_i}\cdots\mO_{A_{N-1}})\nn \\
&&=\Tr(\rho_0 \mO_{A_0}\mO_{A_1}(t_1)\cdots\mO_{A_i}(t_i)\cdots\mO_{A_{N-1}}(t_{N-1})).
\eea
$T_{A_0 A_1 \cdots A_i \cdots A_{N-1}}$ contains only the correlations within the subsystems $A$. We refer to $T_{A_0 A_1 \cdots A_i \cdots A_{N-1}}$ as the \textit{reduced spacetime density matrix} for the combined subsystem $A_0 \cup A_1 \cup \cdots \cup A_{N-1}$ in spacetime.

It can also be shown that if all the subsystems $A_i$ are causally disconnected, the reduced transition operator $T_{A_0 A_1 \cdots A_i \cdots A_{N-1}}$ is Hermitian, and thus reduces to the usual density matrices. However, if any two subsystems, say $A_i$ and $A_j$, are causally connected, $T_{A_0 A_1 \cdots A_i \cdots A_{N-1}}$ is generally non-Hermitian, so that its spectrum can be complex-valued\footnote{In Section~\ref{Section_reduced_moments_universal}, we will show that the exact form of $T_{A_0 A_1 \cdots A_i \cdots A_{N-1}}$ depends on the choice of subsystems as well as the Hamiltonian or evolution operator. The coupling between subsystems is important. For two causally connected subsystems, if there is no interaction between them, the reduced spacetime density matrix can still be Hermitian. If interactions exist, the operator is generally expected to be non-Hermitian.}.

With the reduced spacetime density matrix, one can define entropy-related quantities that reflect the entanglement or correlations between these general spacetime subregions $A_i$. This can be viewed as a straightforward generalization of the entanglement measures for spacelike-separated subregions to the timelike case. The von Neumann entropy is a natural choice. One can define the R\'enyi entropy for $T_{A_0 A_1 \cdots A_i \cdots A_{N-1}}$ as
\bea\label{Renyi}
S_n(T_{A_0 A_1 \cdots A_i \cdots A_{N-1}}) = \frac{\log \tr T_{A_0 A_1 \cdots A_i \cdots A_{N-1}}^n}{1-n},
\eea
for integer $n$, and the von Neumann entropy is given by
\bea
S(T_{A_0 A_1 \cdots A_i \cdots A_{N-1}}) = \lim_{n \to 1} S_n(T_{A_0 A_1 \cdots A_i \cdots A_{N-1}}).
\eea

If all the subsystems $A_i$ are causally disconnected, the above definitions reduce to the usual R\'enyi entropy and entanglement entropy, which are positive-valued. However, if any two subsystems are causally connected, the resulting entropies can generally be complex-valued. In this case, they can be interpreted as pseudoentropy, as proposed in \cite{Nakata:2020luh}.

In QFTs, one can use the replica method to evaluate these quantities. To do so, it is necessary to understand how to represent the reduced spacetime density matrix in terms of the path integral. This is a straightforward generalization of the Schwinger–Keldysh representation of $T_{C_0 C_1 \cdots C_i \cdots C_{N-1}}$, as discussed in Sections~\ref{section_SK} and \ref{section_GMT}. It also generalizes the approach in \cite{Dong:2016hjy}, where the author considered the Schwinger–Keldysh representation of $\rho(t)$ with the motivation of proving the HRT formula.

\begin{figure}[htbp]
  \centering
  \includegraphics[width=0.9\textwidth]{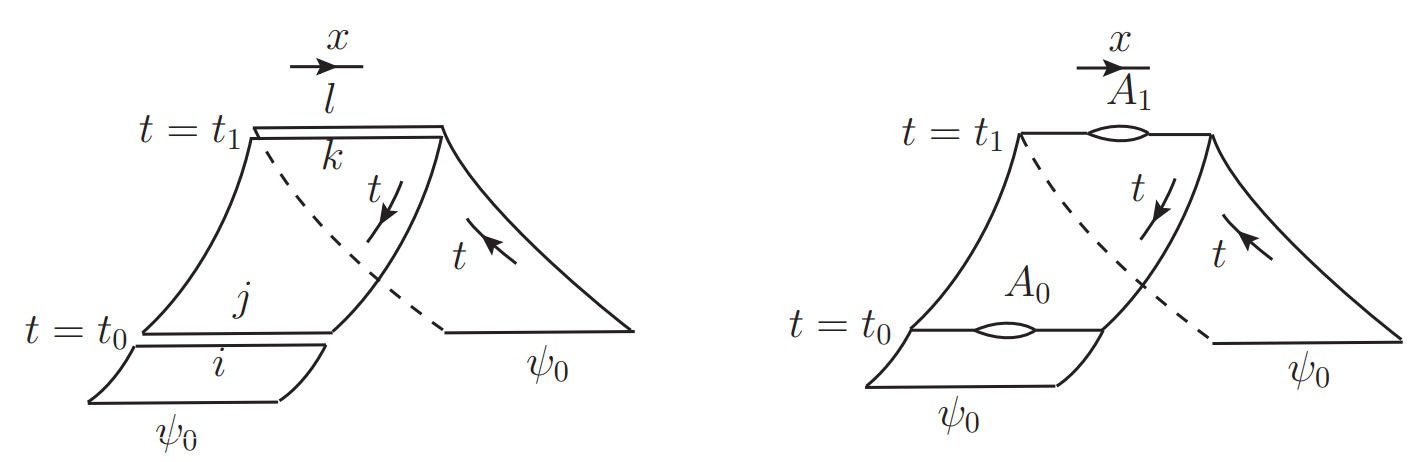}
  \caption{(left) Schwinger–Keldysh representation of the operator $T_{C_0 C_1}$ including the spatial directions. The arrow along $t$ indicates the direction of time evolution. (right) Schwinger–Keldysh representation of the reduced spacetime density matrix $T_{A_0 A_1}$. The partial trace over $\bar A_0$ and $\bar A_1$ is implemented by gluing the boundary conditions on these complementary subregions, while cuts remain at the subregions $A_0$ and $A_1$.
}\label{Fig_SK_T01_reduced}
\end{figure}

For simplicity, we focus on the reduced transition operator $T_{A_0 A_1}$. To illustrate the partial trace over $\bar A_0$ and $\bar A_1$, it is convenient to include the spatial directions in the Schwinger–Keldysh formalism. In Fig.~\ref{Fig_SK_T01_reduced}, we show the representations of $T_{C_0 C_1}$ and $T_{A_0 A_1}$ with the spatial directions included.

With this representation, one can use the real-time replica method to evaluate the R\'enyi entropy of $T_{A_0 A_1}$. Some examples are discussed in \cite{Milekhin:2025ycm} and \cite{Gong:2025pnu}. We refer the reader to these references for more details on the calculations.

\subsection{Super-operator}\label{Section_super}
In this section, we introduce a super-operator $\mathcal{T}$ acting on the density matrix. We will show that the transition operator $T_{C_0 C_1}$ can be reformulated as a super-operator acting on the initial density matrix $\rho_0$. Using this super-operator formalism, one can also obtain generalizations of the spacetime density matrix. In this section, we set $t_1 = t$.

Motivated by the expression (\ref{Two_Form1}) of $T_{C_0 C_1}$, we introduce the super-operator $\mathcal{T}$ acting on operators in the Hilbert space $\mathcal{H}_0 \otimes \mathcal{H}_1$ as
\bea
\mathcal{T}(O_0 \otimes O_1) := U^\dagger |i\rangle \langle j| U O_0 \otimes |j\rangle \langle i| O_1,
\eea
where $U := U(t,t_0)$, and $O_0$ and $O_1$ denote operators in the Hilbert spaces $\mathcal{H}_0$ and $\mathcal{H}_1$, respectively.

\begin{figure}[htbp]
  \centering
  \includegraphics[width=0.23\textwidth]{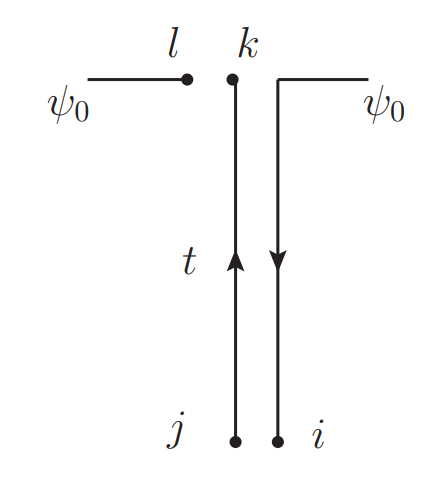}
  \caption{Schwinger-Keldysh representation of the dual transition operator $\tilde{T}_{C_0C_1}$
}\label{Fig_SK_T01_dual}
\end{figure}

With this definition, the operator $T_{C_0 C_1}$ can be expressed as
\bea
T_{C_0 C_1} = \mathcal{T}(\rho_0 \otimes I).
\eea
With the super-operator, it is natural to define a dual operator for $T_{C_0 C_1}$, denoted as
\bea\label{dual_operator}
\tilde{T}_{C_0 C_1} := \mathcal{T}(I \otimes \rho_0) = U^\dagger |l\rangle \langle k| U \otimes |k\rangle \langle l| \rho_0.
\eea
One can check that
\bea
tr(\tilde{T}_{C_0 C_1} \mOz \mOo) = \langle k| U  \mOz U^\dagger |l\rangle  \langle l| \rho_0 \mOo |k\rangle = \Tr(\rho_0 \mOo \mOz(t_0 - t)),
\eea
which gives the time-ordered correlation function, but with the time evolution of $\mOz$ running backward. We also have the relations
\bea\label{dual_operator_partial}
tr_{C_0} \tilde{T}_{C_0 C_1} = \rho_0 \quad tr_{C_1} \tilde{T}_{C_0 C_1} = U^\dagger \rho_0 U = \rho(t_0 - t).
\eea
This means $\tilde{T}_{C_0C_1}$ includes the correlation information on Cauchy surfaces at $t_0$ and $t_0-t$.
The dual operator can be written as
\bea
&&\tilde{T}_{C_0C_1}=\langle i|U^\dagger|l\rangle \langle k|U|j\rangle \langle l|\rho_0|m\rangle |i\rangle\langle j|\otimes   |k\rangle\langle m|\nn \\
&&\phantom{\tilde{T}_{C_0C_1}}=\langle i|U^\dagger\rho_0|l\rangle \langle k|U|j\rangle  |i\rangle\langle j|\otimes   |k\rangle\langle l|.
\eea
Assume the pure density matrix $\rho_0=|\psi_0\rangle \langle \psi_0|$. We can express the dual operator $\tilde{T}_{C_0C_1}$ via the path integral shown in Fig.~\ref{Fig_SK_T01_dual}.

The dual operator $\tilde{T}_{C_0 C_1}$ appears to play a similar role as $T_{C_0 C_1}$. As we will see in the following section, the dual operator satisfies a different, and in fact simpler, equation of motion compared to $T_{C_0 C_1}$.

\subsection{Liouville–von Neumann equation for spacetime density matrix}

Let us consider a closed quantum system with a time-independent Hamiltonian $H$. The evolution operator is given by $U(t,t_0) = e^{-iH(t-t_0)}$. The density matrix $\rho(t) = U(t,t_0) \rho_0 U(t,t_0)^\dagger$ then satisfies the Liouville–von Neumann equation
\bea\label{LN_densitymatrix}
i \frac{\partial \rho(t)}{\partial t} = [H, \rho(t)].
\eea
We now consider a generalization of the density matrix $\rho(t)$. Accordingly, the operator $T_{C_0 C_1}$ and its dual operator $\tilde{T}_{C_0 C_1}$ also satisfy certain equations of motion. Here we set $t_1 = t$. By using the expression (\ref{Two_Form1})\footnote{It is also possible to use (\ref{Two_Form2}) to derive the equation of motion. The result is similar, so we do not repeat the derivation here.}, the derivation is straightforward, yielding
\bea
&&\frac{\partial T_{C_0C_1}}{\partial t}=i H U^\dagger|k\rangle \langle l| U \rho_0 \otimes |l\rangle \langle k|-iU^\dagger|k\rangle \langle l| U H \rho_0 \otimes |l\rangle \langle k|\nn \\
&&\phantom{\frac{\partial T_{C_0C_1}}{\partial t}}=i [H\otimes I, T_{C_0C_1}]+iU^\dagger|k\rangle \langle l| U \rho_0 H\otimes |l\rangle \langle k|-iU^\dagger|k\rangle \langle l| U H \rho_0 \otimes |l\rangle \langle k|\nn \\
&&\phantom{\frac{\partial T_{C_0C_1}}{\partial t}}=i [H\otimes I, T_{C_0C_1}]+iU^\dagger|k\rangle \langle l| U [\rho_0,H]\otimes |l\rangle \langle k|\nn \\
&&\phantom{\frac{\partial T_{C_0C_1}}{\partial t}}=i [H\otimes I, T_{C_0C_1}]+U^\dagger|k\rangle \langle l| U \frac{\partial\rho_0}{\partial t}\otimes |l\rangle \langle k|,
\eea
where $\frac{\partial\rho_0}{\partial t}$ should be understood as $\frac{\partial \rho(t)}{\partial t}|_{t=t_0}$. By using the super-operator $\mathcal{T}$, the equation of motion can be written in a more compact form as
\bea\label{LN_transition}
\frac{\partial\mathcal{T}(\rho_0\otimes I)}{\partial t}=i [H\otimes I, \mathcal{T}(\rho_0\otimes I)]+\mathcal{T}(\frac{\partial \rho_0}{\partial t}\otimes I).
\eea
The form of the equation of motion is slightly different from the Liouville–von Neumann equation for the density matrix (\ref{LN_densitymatrix}). The last term in Eq.~\ref{LN_transition} depends only on the initial ``velocity'' of the density matrix, $\frac{\partial \rho_0}{\partial t}$, which is therefore time-independent. The equation of motion (\ref{LN_transition}) remains linear, which means that for $\rho_m = \sum_i p_i \rho_i$, the transition operator $\mathcal{T}(\rho_m \otimes I)$ also satisfies the equation of motion (\ref{LN_transition}) if each $\rho_i$ does. This follows from the fact that the super-operator $\mathcal{T}$ is a linear mapping.

We can now also derive the equation of motion for the dual operator $\tilde{T}_{C_0 C_1}$. Using (\ref{dual_operator}), we have
\bea
&&\frac{\partial \tilde{T}_{C_0C_1}}{\partial t}=iH U^\dagger|i\rangle \langle j|U\otimes |j\rangle \langle i|\rho_0-i U^\dagger|i\rangle \langle j|UH\otimes |j\rangle \langle i|\rho_0\nn \\
&&\phantom{\frac{\partial \tilde{T}_{C_0C_1}}{\partial t}}=i[H\otimes I, \tilde{T}_{C_0C_1}].
\eea
It is also useful to express the equation of motion in terms of the super-operator; the result is
\bea
\frac{\partial \mathcal{T}(I\otimes \rho_0)}{\partial t}=i[H\otimes I, \mathcal{T}(I\otimes \rho_0)].
\eea
The equation of motion for the dual operator is much simpler than that of the transition operator, as it does not include the last time-independent term. As a check of the above equation, we can take the partial trace over $C_1$. Using (\ref{dual_operator_partial}), we have
\bea
\frac{\partial tr_{C_1} \tilde{T}_{C_0C_1}}{\partial t}=\frac{\partial \rho(t_0-t)}{\partial t}=i [H,\rho(t_0-t)],
\eea
which is Liouville-von Neumann equation for density matrix with backward time evolution.

\subsection{Generalization of spacetime density matrix}\label{Section_generalization}
The transition operator $T_{C_0C_1}$ and its generalization to multiple timeslices, $T_{C_0C_1\cdots C_i\cdots C_{N-1}}$, are defined with the motivation of reproducing time-ordered or anti-time-ordered correlation functions. This naturally leads to the Schwinger-Keldysh path integral with cuts on different time slices along the forward or backward time evolution branches. From the perspective of the super-operator $\mathcal{T}$ introduced in Section~\ref{Section_super}, the transition operator $T_{C_0C_1}$ is a special case given by $\mathcal{T}(\rho_0 \otimes I)$. In this section, we introduce some generalizations of the transition operator, which can be naturally defined and understood either in the path integral formalism or from the super-operator viewpoint.

One natural generalization is motivated by the Schwinger-Keldysh representation of $T_{C_0C_1}$ or $T_{C_0C_1\cdots C_i\cdots C_{N-1}}$, as shown in Fig.\ref{Fig_SK_T01} and Fig.\ref{Fig_SK_T0N}. In these representations, the initial and final states of the path integral are the same, $|\psi_0\rangle$, and the transition operators involve cuts only on one branch of the path integral. A natural generalization is to fix the initial state as $|\phi_0\rangle$, while allowing the final state to be state $|\psi_0\rangle$. In general, during the time evolution of a system, the final state differs from the initial state. In some contexts, the final state is fixed. For example, in quantum information theory, weak values involve post-selected states \cite{Aharonov:1988xu}\cite{Dressel}. In scattering processes, one is generally interested in specific final states, which are usually different from the initial state. Motivated by these considerations, we introduce a generalization of the transition operator $T_{C_0C_1}$.

\begin{figure}[htbp]
  \centering
  \includegraphics[width=0.23\textwidth]{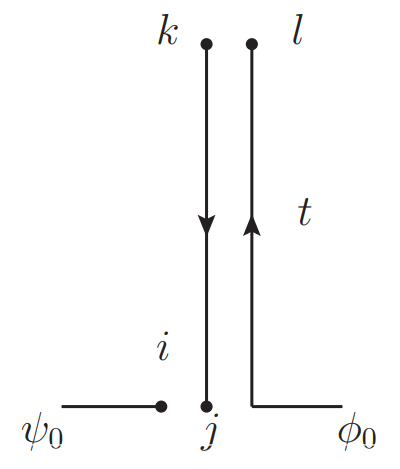}
  \caption{Schwinger-Keldysh representation of the transition operator $\mathcal{F}_{C_0C_1}$ with an initial state $|\phi_0\rangle$ and final state $|\psi_0\rangle$.
}\label{Fig_SK_T01_g2}
\end{figure}

It is useful to illustrate the path integral representation of this generalized transition operator, which is shown in Fig.~\ref{Fig_SK_T01_g2}.
 According to the path integral we can write down the transition operator for this case:
 \bea
 \mathcal{F}_{C_0C_1}=\langle l|U| \phi_0\rangle \langle \psi_0 |i\rangle \langle j|U^\dagger|k\rangle |j\rangle \langle i|\otimes |l\rangle \langle k|.
\eea
The trace of $\mathcal{F}_{C_0C_1}$ is given by
\bea
tr_{C_0C_1}\mathcal{F}_{C_0C_1}=\langle \psi_0|\phi_0\rangle,
\eea
which is the transition amptitute from $|\phi_0\rangle$ to $|\psi_0\rangle$.
Assume $\langle \phi_0|\psi_0\rangle\ne 0$. It is also possible to define the normalized version:
\bea
F_{C_0C_1}=\frac{\mathcal{F}_{C_0C_1}}{\langle \psi_0|\phi_0\rangle}.
\eea
By using the super-operator we can write $F_{C_0C_1}$ as
\bea
F_{C_0C_1}=\mathcal{T}(T^{\phi_0|\psi_0}\otimes I),
\eea
where $T^{\phi_0|\psi_0}$ is the transition matrix defined in \cite{Nakata:2020luh},
\bea
T^{\phi_0|\psi_0}:=\frac{|\phi_0\rangle \langle \psi_0|}{\langle \psi_0|\phi_0\rangle}.
\eea
We can identify the following properties of $F_{C_0C_1}$:
\bea
&&tr_{C_1}F_{C_0C_1} =T^{\phi_0|\psi_0},\quad tr_{C_0}F_{C_0C_1}= U T^{\phi_0|\psi_0} U^\dagger,\nn \\
&&tr_{C_0C_1}(F_{C_0C_1}\mOz \mOo)=tr (T^{\phi_0|\psi_0}\mOz \mOo(t_1))=\frac{\langle \psi_0| \mOz \mOo(t_1)|\phi_0\rangle}{\langle \psi_0|\phi_0\rangle}.
\eea
It is also possible to define the Hermtian conjugation of $F_{C_0C_1}$, which satisfies
\bea
tr(F_{C_0C_1}^\dagger\mOz \mOo)=\frac{\langle \phi_0| \mOo(t_1)\mOz|\psi_0\rangle}{\langle \phi_0|\psi_0\rangle}.
\eea
We have 
\bea
\mathcal{F}_{C_0C_1}^\dagger=\langle k |U|j\rangle \langle i|\psi_0\rangle \langle \phi_0|U^\dagger |l\rangle |i\rangle\langle j|\otimes |k\rangle\langle l|,
\eea
and the normalized operator $F_{C_0C_1}^\dagger=\frac{\mathcal{F}_{C_0C_1}^\dagger}{\langle \phi_0|\psi_0\rangle}$.
We can also define the dual operator 
\bea
\tilde{F}_{C_0C_1}:=\mathcal{T}(I\otimes T^{\phi_0|\psi_0})=U^\dagger|l\rangle \langle k|U\otimes |k\rangle \langle l|T^{\phi_0|\psi_0},
\eea
which can also give the anti-time-ordered correlator.
\bea
tr_{C_0C_1}(\tilde{F}_{C_0C_1}\mOz \mOo)= tr(T^{\phi_0|\psi_0}\mOo U\mOz U^\dagger)=tr(T^{\phi_0|\psi_0}\mOo \mOz(t_0-t) ).
\eea
\begin{figure}[htbp]
  \centering
  \includegraphics[width=0.6\textwidth]{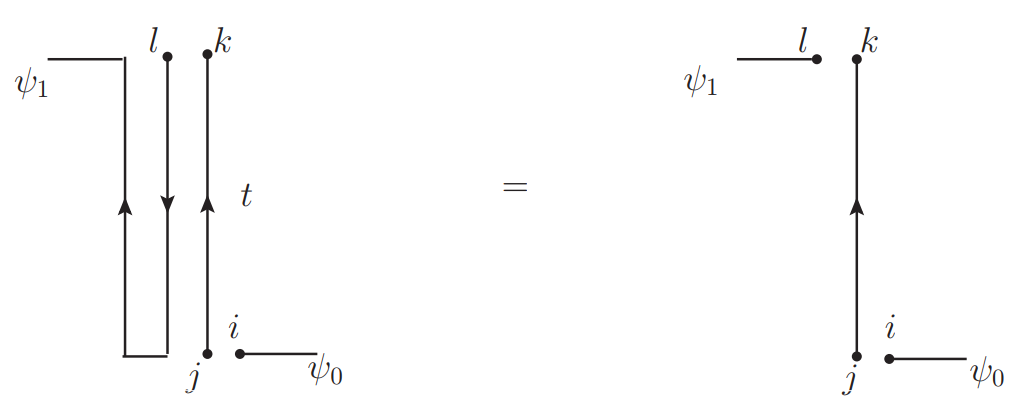}
  \caption{Path integral representation of the transition operator $\mathcal{F}_{C_0C_1}^\dagger$ with a final state $|\phi_0\rangle=U^\dagger|\psi_1\rangle$.
}\label{Fig_SK_T01_g3}
\end{figure}
A special case of $\mathcal{F}_{C_0C_1}^\dagger$ is the state $|\phi_0\rangle = U^\dagger |\psi_1\rangle$, where $|\psi_1\rangle$ is arbitrary final state. The path integral representation is shown in Fig.~\ref{Fig_SK_T01_g3}. The unnormalized transition operator in this special can be expressed as
\bea\label{Generalization_F}
 \mathcal{F}_{C_0C_1,s}^\dagger=\langle \psi_1| l\rangle \langle i |\psi_0\rangle \langle k|U|j\rangle |i\rangle \langle j|\otimes |k\rangle \langle l|.
\eea
The normalization of the above operator is given by
\bea
tr \mathcal{F}_{C_0C_1,s}^\dagger=\langle \psi_1| U|\psi_0\rangle,
\eea
which is just the time-evolution amplitude from the initial state $|\psi_0\rangle$ to $|\psi_1\rangle$. This indicates the operator $\mathcal{F}_{C_0C_1,s}^\dagger$ is closely related to amplitude of the dynamical evlolution. Specificly, in the scattering process, $|\psi_0\rangle$ and $|\psi_1\rangle$ can be chosen as the in-state with $t_0\to -\infty$ and out-state with $t_1\to +\infty$. $\mathcal{F}_{C_0C_1,s}^\dagger$ do inlcude the S-matrix information of the theory.

\begin{figure}[htbp]
  \centering
  \includegraphics[width=0.23\textwidth]{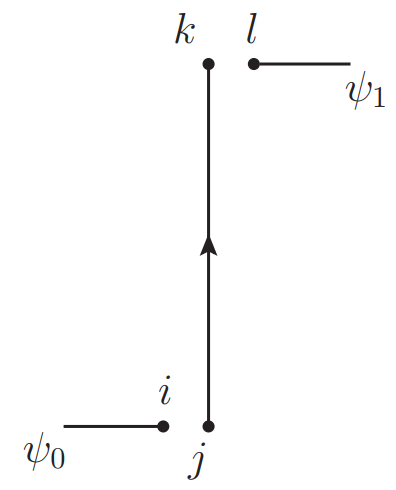}
  \caption{Path integral representation of the transition operator $\mathcal{F}_{C_0C_1}$ with a final state $|\phi_0\rangle=U^\dagger|\psi_1\rangle$.
}\label{Fig_SK_T01_g3_HJ}
\end{figure}

We can also define the operator $\mathcal{F}_{C_0C_1,s}$, the path integral representation is given by Fig.~(\ref{Fig_SK_T01_g3_HJ}).
We are particularly interested in studying the entanglement between \textit{subregions} defined on the Cauchy surfaces $t=t_0$ and $t=t_1$. For $\mathcal{F}_{C_0C_1}$ and $\mathcal{F}_{C_0C_1}^\dagger$, it is also straightforward to define the corresponding reduced operators. In the Section.~\ref{Section_example}, we will illustrate this idea through explicit examples.

Another generalization, involving two density matrices, is presented in Appendix~\ref{Section_another}. This construction may be related to the ``trace distance'' between the two density matrices. In certain cases, it can also be connected to the operator $\mathcal{F}_{C_0C_1}$.

\section{Properties of spacetime density matrix}

In this section, we discuss the general properties of the spacetime density matrix and the reduced operators.

\subsection{Moments of spacetime density matrix}\label{Section_moment_transition operator}

The transition operator $T_{C_0C_1...C_i...C_{N-1}}$ contains the information of correlation functions across the $N$ time slices. From the expression (\ref{mutiple_transition}), these operators are generally time-dependent. In this section, we study the moments of the transition operators:
\bea
tr T_{C_0C_1...C_i...C_{N-1}}^n,\quad tr(T_{C_0C_1...C_i...C_{N-1}}^\dagger)^n,\quad tr(T_{C_0C_1...C_i...C_{N-1}}T_{C_0C_1...C_i...C_{N-1}}^\dagger)^m,
\eea
where $n$ and $m$ are integers. 

Let us first consider $T_{C_0C_1}$. Using the expression (\ref{Two_Form1}), one can show that
\bea\label{trace_T01_2moment}
&&trT_{C_0C_1}^2= (Tr\rho_0)^2, \quad tr(T_{C_0C_1}^\dagger)^2= (Tr\rho_0)^2,\nn \\
&&tr (T_{C_0C_1}T_{C_0C_1}^\dagger)=tr (T_{C_0C_1}^\dagger T_{C_0C_1})=d \ Tr \rho_0^2,
\eea
where $d$ is the dimension of the Hilbert space. These quantities depend only on the state $\rho_0$ and the dimension of the Hilbert space, but are independent of time. For an infinite-dimensional Hilbert space, the quantity $tr(T_{C_0C_1}T_{C_0C_1}^\dagger)$ is not well-defined without regularization.

\begin{figure}[htbp]
  \centering
  \includegraphics[width=0.7\textwidth]{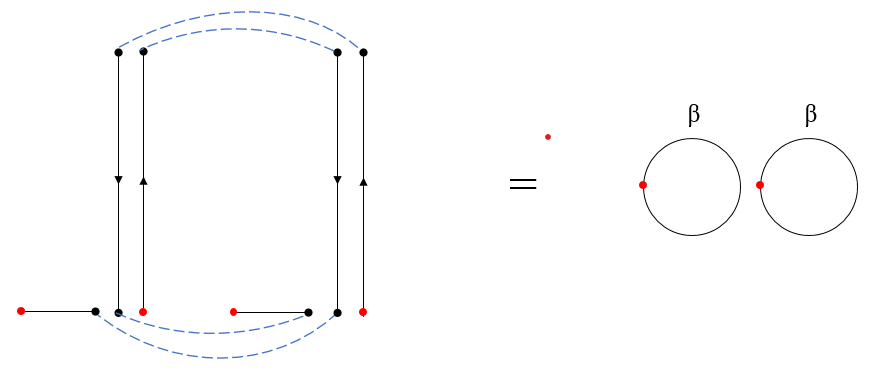}
  \caption{Path integral representation $trT_{C_0C_1}^2$. The dash line means the identification. The circle on the right hand side of equation refer to the path integral on imaginary time direction, the circumference is $\beta$.
}\label{Fig_trace_T01_2moment}
\end{figure}

\begin{figure}[htbp]
  \centering
  \includegraphics[width=0.7\textwidth]{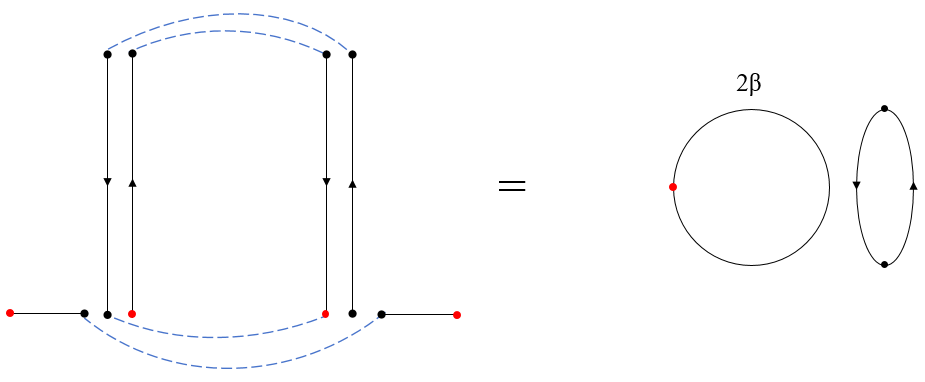}
  \caption{Path integral representation $tr(T_{C_0C_1}T_{C_0C_1}^\dagger)$. The dash line means the identification. The circle on the right hand side of equation refer to the path integral on imaginary time direction, the circumference is $2\beta$, while the loop means the path integral on the real time direction, which gives the dimension of Hilbert space.
}\label{Fig_trace_T01_2moment_TTHJ}
\end{figure}

The above results can be understood using the path integral representations of $T_{C_0C_1}$ and $T_{C_0C_1}^\dagger$. Here we assume the initial state is $\rho_0=e^{-\beta H}$. The corresponding representations of $T_{C_0C_1}$ and $T_{C_0C_1}^\dagger$ are shown in Fig.\ref{Fig_SK_T01_thermal}. The results (\ref{trace_T01_2moment}) are illustrated in Fig.\ref{Fig_trace_T01_2moment} and Fig.~\ref{Fig_trace_T01_2moment_TTHJ}, and can be straightforwardly generalized to an arbitrary initial state $\rho_0$.


In Appendix~\ref{Section_moments_any_slices}, using the path integral representation, we provide more examples of evaluating the moments of $T_{C_0C_1}$ and $T_{C_0C_1}^\dagger$. We expect the following results:
\bea\label{n_moment}
&&tr(T_{C_0C_1}^n)=\begin{cases}
Tr\rho_0^n, & n \in \text{odd},  \\
(Tr \rho_0^{n/2})^2,  & n \in  \text{even}.
\end{cases}
\eea
\bea\label{singular_value}
tr(T_{C_0C_1}T_{C_0C_1}^\dagger)^n=d \ Tr \rho_0^{2n}.
\eea
The results indicate that the moments of $T_{C_0C_1}$ and $T_{C_0C_1}^\dagger$ are time-independent, depending only on the initial density matrix $\rho_0$ and the dimension of the Hilbert space. This suggests that the spacetime density matrix $T_{C_0C_1}$ belongs to a special class of operators in finite-dimensional Hilbert spaces. If $T_{C_0C_1}$ is diagonalizable, the above relations can be used to infer certain properties of its eigenvalues.

A first observation is that $T_{C_0C_1}$ and $T_{C_0C_1}^\dagger$ share the same eigenvalues, even though $T_{C_0C_1}$ is generally non-Hermitian. Since all the $n$-th moments of $T_{C_0C_1}$ are positive, its eigenvalues must either be real or appear in complex-conjugate pairs\footnote{In Section.~\ref{Section_example} we  study $T_{C_0C_1}$ for a two-qubit system. In that example, the eigenvalues of $T_{C_0C_1}$ appear to be all real, but can take negative values.}. This indicates that $T_{C_0C_1}$ is a pseudo-Hermitian operator, following a theorem proved in \cite{Mostafazadeh:2001jk}, see also the applications of pseudo-Hermiticity in the context of pseudoentropy discussed in \cite{Guo:2022jzs}.

For a non-Hermitian matrix $T_{C_0C_1}$, one can also perform a singular value decomposition (SVD)\footnote{See \cite{Parzygnat:2023avh} for applications of SVD of the transition matrix $T^{\phi|\psi}$ in defining the so-called SVD entropy.}.
\bea
T_{C_0C_1}=u \Sigma v^\dagger,
\eea
where $u$ and $v$ are unitary matrices, $\Sigma=\text{diag}(\sigma_1,\cdots,\sigma_{r},0,\cdots,0)$, $r$ is rank of $T_{C_0C_1}$ and $\sigma_i$ ($i=1,\cdots,r$) are the singular values of $T_{C_0C_1}$. By taking $n\to 1/2$ in (\ref{singular_value}) and assuming $Tr\rho_0=1$  we have
\bea
\sum_i^{r}\sigma_i=d.
\eea
This would give a constraint on the eigenvalues $\{\lambda_i\}$ ($i=1,\cdots,d^2$) of $T_{C_0C_1}$ that
\bea
\sum_{i}^{d^2}|\lambda_i|\le \sum_i^{r}\sigma_i=d.
\eea
In the above, we discussed some basic properties of the eigenvalues of $T_{C_0C_1}$ based on the relations (\ref{n_moment}) and (\ref{singular_value}). In fact, one can, in principle, reconstruct the eigenvalues of $T_{C_0C_1}$ if all the $n$-th moments of $\rho_0$ are known. This suggests that there exists a direct relation between the eigenvalues of $T_{C_0C_1}$ and those of $\rho_0$.

\begin{figure}[htbp]
  \centering
  \includegraphics[width=0.7\textwidth]{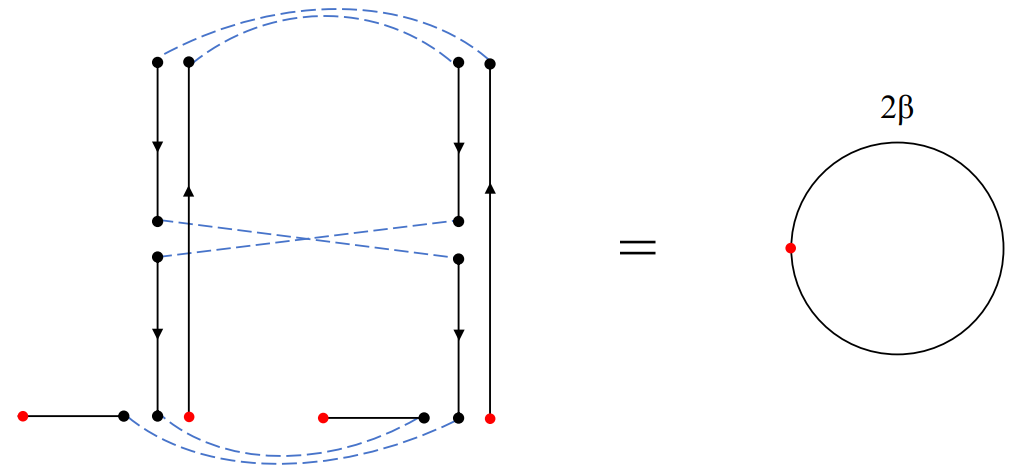}
  \caption{Path integral representation $trT_{C_0C_1C_2}^2$. 
}\label{Fig_trace_T012_2moment}
\end{figure}

The above results can also be generalized to the case of multiple time slices, $T_{C_0C_1\cdots C_i \cdots C_{N-1}}$. One can evaluate the moments using the explicit expression (\ref{mutiple_transition}). However, it is often more convenient to use the path integral representation for this purpose. As an example, let us consider $T_{C_0C_1C_2}$. The result is:
\bea
&&tr(T_{C_0C_1C_2}^2)= \Tr\rho_0^2,\nn\\
&&tr(T_{C_0C_1C_2}T_{C_0C_1C_2}^\dagger)= d\ \Tr\rho_0^2.
\eea
The path integral evaluation of the above results for the thermal state is illustrated in Fig.~\ref{Fig_trace_T012_2moment}.
We can also establish the following result:
\bea\label{2moment_any_slices}
&&tr(T_{C_0C_1\cdots C_i\cdots C_{N-1}}^2)= \begin{cases}
Tr\rho_0^2, & N \in \text{odd},  \\
(Tr \rho_0)^2,  & N \in  \text{even}.
\end{cases}\\
&&tr(T_{C_0C_1\cdots C_i\cdots C_{N-1}}T_{C_0C_1\cdots C_i\cdots C_{N-1}}^\dagger)= d\ \Tr\rho_0^2.
\eea
Some explicit examples are provided in Appendix~\ref{Section_moments_any_slices}.

\subsection{Reduced spacetime density matrix and its moments}\label{Section_reduced_moments_universal}
In Section~\ref{Section_reduced_transition}, we showed how to define the reduced transition operator and associated entropy-related quantities, which capture the entanglement between general spacetime subregions. This provides a foundation for defining and evaluating entanglement entropy for causally connected subregions. In this section, we study general properties of the reduced transition operator. While the moments of the full transition operator are time-independent, the moments of the reduced transition operator 
$T_{A_0A_1\cdots A_i\cdots A_{N-1}}$ are sensitive to time evolution. This makes the entropy-related quantities valuable indicators of the system's dynamics.

Let us briefly describe our basic setup. We focus first on the transition operator $T_{C_0C_1}$. The systems at times $t=t_0$ and $t=t_1$ are divided into subsystems $A_0, \bar A_0$ and $A_1, \bar A_1$, respectively. For instance, in a two-qubit system, $A_0$ and $A_1$ could represent one qubit at times $t_0$ and $t_1$. In QFTs, $A_0$ and $A_1$ could correspond to spatial subregions on the Cauchy surfaces $t_0$ and $t_1$.

We denote the bases of $\mathcal{H}_0$ and $\mathcal{H}_1$ as $\{|I\rangle \otimes |\bar J\rangle\}$, where $|I\rangle$ and $|\bar J\rangle$ are bases for $A_{0(1)}$ and $\bar A_{0(1)}$, respectively.\footnote{Here we assume discrete eigenbases. The notation can be extended to continuous cases. One can consider $A$ and $\bar A$ as subsystems with Hamiltonians $H_A$ and $H_{\bar A}$, possibly interacting via $H_{A\bar A}$. Alternatively, $A \cup \bar A$ can be treated as the full system with Hamiltonian $H_{A\bar A}$, with $A$ and $\bar A$ as subsystems. Note that in QFTs, the operator algebra differs from the finite-dimensional case, so care is required when extending these results to field theories.} With this notation, the transition operator $T_{C_0C_1}$ can be written as
\bea
T_{C_0C_1}=\langle L| \langle \bar L |U \rho_0 |I\rangle |\bar I\rangle  \langle J| \langle \bar J|U^\dagger |K\rangle |\bar K\rangle  |J\rangle |\bar J\rangle \langle I| \langle \bar I| \otimes |L\rangle |\bar L\rangle \langle K| \langle \bar K|. 
\eea
Now we have four subsystems $A_0$, $\bar A_0$ and $A_1$, $\bar A_1$. We are interested in the following reduced transition operators\footnote{One can also define other reduced transition operators. Some of them reduce to standard density matrices; for example, 
$tr_{C_0\bar A_1}T_{C_0C_1} = tr_{\bar A_1} \rho(t)$ and 
$tr_{C_1\bar A_0}T_{C_0C_1} = tr_{\bar A_0} \rho_0$. 
Others are non-trivial, e.g., $T_{C_0A_1} := tr_{\bar A_1}T_{C_0C_1}$ or $T_{C_1A_0} := tr_{\bar A_0}T_{C_0C_1}$, and may have interesting applications in certain situations. 
In this paper, however, we will not discuss the properties of these operators.}:
\bea
&&T_{A_0A_1}:=tr_{\bar A_0\bar A_1} T_{C_0C_1}, \quad T_{A_0 \bar A_1}:=tr_{\bar A_0  A_1} T_{C_0C_1},\nn \\
&&T_{\bar A_0 A_1}:= tr_{A_0\bar A_1} T_{C_0C_1},\quad T_{\bar A_0 \bar A_1}=tr_{A_0A_1} T_{C_0C_1}.
\eea 
$T_{A_0A_1}$ and $T_{\bar A_0 \bar A_1}$ share similar properties, while $T_{A_0 \bar A_1}$ and $T_{\bar A_0 A_1}$ share similar properties. Therefore, we will mainly focus on $T_{A_0A_1}$ and $T_{A_0\bar A_1}$. 
We are interested in the R\'enyi entropy (\ref{Renyi}) for these operators. 
To this end, we should evaluate the moments 
\bea
tr \, T_{A_0A_1}^n, \quad tr \, T_{A_0\bar A_1}^n,
\eea
which generally depend on the time evolution and can provide useful information about the dynamics of the system.

It is straigforward to obtain 
\bea\label{reduced_two_party}
&&T_{A_0A_1}=\langle L |\langle \bar K|U\rho_0|I\rangle|\bar I\rangle \langle J| \langle \bar I|U^\dagger |K\rangle |\bar K\rangle |J\rangle\langle I| \otimes |L\rangle \langle K|,\\
&&T_{A_0\bar A_1}=\langle K |\langle \bar L|U\rho_0|I\rangle|\bar I\rangle \langle J| \langle \bar I|U^\dagger |K\rangle |\bar K\rangle |J\rangle\langle I| \otimes |\bar L\rangle \langle \bar K|.
\eea
Now we are ready to study the moments of these reduced spacetime density matrices.

\subsection{Universal properties at short time}
Firstly, we consider the properties of the 2-moments at the short-time limit:
\bea
&&tr(T_{A_0A_1}^2),\quad tr(T_{A_0A_1}T_{A_0A_1}^\dagger),\nn\\
&&tr(T_{A_0\bar A_1}^2),\quad tr(T_{A_0\bar A_1}T_{A_0\bar A_1}^\dagger).
\eea
We set the short-time interval as $t_1-t_0=\delta t$. The evolution operator can be expanded as
\bea
U=e^{-iH\delta t} = 1 - iH \delta t - \frac{1}{2}H^2 \delta t^2 + \cdots.
\eea
This expansion is expected to be valid if $\delta t \ll \langle H \rangle_{\rho_0}$. In general, the Hamiltonian $H$ can be decomposed as
\bea\label{Hamiltonian}
H = H_{A_0} + H_{\bar A_0} + \lambda V,
\eea
where $H_{A_0}$ and $H_{\bar A_0}$ are the Hamiltonians of the subsystems $A_0$ and $\bar A_0$, respectively, $V$ denotes the interaction term, and $\lambda$ is a dimensionless coupling constant. For later purposes, we define the free part of the Hamiltonian as
\bea\label{free_Hamiltonian}
H_0 := H_{A_0} + H_{\bar A_0}.
\eea
In the studies of open quantum systems, $A_0$ can be regarded as the system, $\bar A_0$ as the environment, and $V$ as the interaction between them.
 
\subsubsection{Perturbation result for moments}\label{section_perturbation_short}
We will consider the short time limit. Thus, we can perturbatively calculate the moments of the reduced transition operator. For $T_{A_0A_1}$, we have the following results:
\bea
T_{A_0A_1}=T^{(0)}_{A_0A_1}+T^{(1)}_{A_0A_1}+T^{(2)}_{A_0A_1}+\cdots,
\eea
with
\bea
&&T^{(0)}_{A_0A_1}=\langle L |\langle \bar I|\rho_0|I\rangle|\bar I\rangle  |J\rangle\langle I| \otimes |L\rangle \langle J|,\nn\\
&&\phantom{T^{(0)}_{A_0A_1}}=\langle L |\rho_{A_0}|I\rangle  |J\rangle\langle I| \otimes |L\rangle \langle J|,\nn \\
&&T^{(1)}_{A_0A_1}=i\delta t \langle L|\langle \bar K|\rho_0|I\rangle |\bar I\rangle \langle J|\langle \bar I|H |K\rangle |\bar K\rangle |J\rangle \langle I|\otimes |L\rangle\langle K|\nn \\
&&\phantom{T^{(1)}_{A_0A_1}=}-i\delta t\langle L|\langle \bar I|H\rho_0|I\rangle |\bar I\rangle |J\rangle\langle I| \otimes |L\rangle\langle J|,\nn \\
&&T^{(2)}_{A_0A_1}=\delta t^2 \langle L|\langle \bar K|H\rho_0|I\rangle |\bar I\rangle \langle J| \langle \bar I|H|K\rangle |\bar K\rangle |J\rangle \langle I|\otimes |L\rangle \langle K|\nn \\
&&\phantom{T^{(2)}_{A_0A_1}=}-\frac{1}{2}\delta t^2 \big( \langle L| \langle \bar I|H^2\rho_0|I\rangle |\bar I\rangle |J\rangle \langle I|\otimes |L\rangle \langle J|\nn \\
&&\phantom{T^{(2)}_{A_0A_1}=} +\langle L|\langle \bar K|\rho_0|I\rangle |\bar I\rangle \langle J|\langle \bar I|H^2|K\rangle |\bar K\rangle |J\rangle \langle I| \otimes |L\rangle\langle K| \big),
\eea
where $\rho_{A_0}:=tr_{\bar A_0}\rho_0$. Note for $T^{(0)}_{A_0A_1}$ we have
\bea
(T^{(0)}_{A_0A_1})^\dagger=T^{(0)}_{A_1A_0},
\eea
where $T^{(0)}_{A_1A_0}$ is defined simlar as $T^{(0)}_{A_0A_1}$ but with exchanging the role of $A_0$ and $A_1$.

For $T_{A_0A_1}$, we have
\bea
T_{A_0\bar A_1}=T^{(0)}_{A_0\bar A_1}+T^{(1)}_{A_0\bar A_1}+T^{(2)}_{A_0\bar A_1}+\cdots,
\eea
with
\bea\label{TA0bA1_perturbation}
&&T^{(0)}_{A_0\bar A_1}=\langle J|\langle \bar L|\rho_0|I\rangle|\bar I\rangle  |J\rangle\langle I| \otimes |\bar L\rangle \langle \bar I|,\nn \\
&&T^{(1)}_{A_0\bar A_1}=i\delta t \langle K|\langle \bar L|\rho_0|I\rangle |\bar I\rangle \langle J|\langle \bar I|H |K\rangle |\bar K\rangle |J\rangle \langle I|\otimes |\bar L\rangle\langle \bar K|\nn \\
&&\phantom{T^{(1)}_{A_0\bar A_1}=}-i\delta t\langle J|\langle \bar L|H\rho_0|I\rangle |\bar I\rangle |J\rangle\langle I| \otimes |\bar L\rangle\langle \bar I|,\nn \\
&&T^{(2)}_{A_0\bar A_1}=\delta t^2 \langle K|\langle \bar L|H\rho_0|I\rangle |\bar I\rangle \langle J| \langle \bar I|H|K\rangle |\bar K\rangle |J\rangle \langle I|\otimes |\bar L\rangle \langle \bar K|\nn \\
&&\phantom{T^{(2)}_{A_0A_1}=}-\frac{\delta t^2}{2}\big( \langle J| \langle \bar L|H^2 \rho_0|I\rangle |\bar I\rangle |J\rangle \langle I|\otimes |\bar L\rangle \langle \bar I|\nn\\
&&\phantom{T^{(2)}_{A_0A_1}=}+\langle K|\langle \bar L|\rho_0|I\rangle |\bar I\rangle \langle J|\langle \bar I| H^2 |K\rangle |\bar K\rangle |J\rangle \langle I|\otimes |\bar L\rangle \langle \bar K|\big).
\eea
The first-order term $T^{(0)}_{A_0\bar A_1}$ is just the operator $\rho_0$ in the basis $|I\rangle |\bar J\rangle$. This is expected, since in the limit $\delta t\to 0$, the systems $C_0$ and $C_1$ are the same. The partial trace over $\bar A_0$ and $A_1$ is equal to the partial trace over $\bar A_0$ and $A_0$ or over $\bar A_1$ and $A_1$, so the result gives the density matrix $\rho_0$.

For the moment $tr T_{A_0A_1}^2$, we have the following results:
\bea
tr T_{A_0A_1}^2=tr(T^{(0)}_{A_0A_1})^2+2tr(T^{(0)}_{A_0A_1}T^{(1)}_{A_0A_1})+tr(T^{(1)}_{A_0A_1})^2+2tr(T^{(0)}_{A_0A_1}T^{(2)}_{A_0A_1})+\cdots,
\eea
where we keep terms only up to order $O(\delta t^2)$. With some calculations, we have
\bea\label{perturbation_short_TA0A1}
&&tr(T^{(0)}_{A_0A_1})^2=(Tr\rho_{A_0})^2=1,\nn\\
&&2tr(T^{(0)}_{A_0A_1}T^{(1)}_{A_0A_1})=-2i\delta t \Tr\left[(\rho_0-\rho_{A_0}\otimes \rho_{\bar A_0})H\right],\nn\\
&&tr (T_{A_0A_1}^{(1)})^2=-\delta t^2Tr_{A_0} \big[ Tr_{\bar A_0}(\rho_0 |J\rangle \langle I|H)Tr_{\bar A_0}(\rho_0|I\rangle \langle J|H)\big],\nn\\
&&\phantom{tr (T_{A_0A_1}^{(1)})^2=}-\delta t^2 [Tr(H\rho_0)]^2+2\delta t^2 Tr_{A_0} \big[Tr_{\bar A_0}(\rho_{\bar A_0}H)Tr_{\bar A_0}(H\rho_0) \big],\nn \\
&&2tr(T^{(0)}_{A_0A_1}T^{(2)}_{A_0A_1})=2\delta t^2 Tr_{\bar A_0}\big[Tr_{A_0}(H\rho_0)Tr_{A_0} (\rho_{A_0}H)\big]\nn \\
&&\phantom{2tr(T^{(0)}_{A_0A_1}T^{(2)}_{A_0A_1})=}-\delta t^2 Tr(H^2 \rho_0)-\delta t^2 Tr[\rho_{A_0}\otimes \rho_{\bar A_0}H^2].
\eea
For the moment $tr T_{A_0\bar A_1}^2$, we would have the following results:
\bea\label{perturbation_short_TA0barA1}
tr T_{A_0\bar A_1}^2=tr(T^{(0)}_{A_0\bar A_1})^2+2tr(T^{(0)}_{A_0\bar A_1}T^{(1)}_{A_0\bar A_1})+tr(T^{(1)}_{A_0\bar A_1})^2+2tr(T^{(0)}_{A_0\bar A_1}T^{(2)}_{A_0\bar A_1})+\cdots,
\eea
with
\bea
&&tr(T^{(0)}_{A_0\bar A_1})^2=\Tr\rho_0^2,\nn \\
&&2tr(T^{(0)}_{A_0\bar A_1}T^{(1)}_{A_0\bar A_1})=2i\delta t\big[\Tr_{\bar A_0}(\Tr_{A_0}(\rho_0|I\rangle \langle J|H)\Tr_{A_0}(\rho_0|J\rangle \langle I|))-Tr(H\rho_0^2) \big],\nn \\
&&tr(T^{(1)}_{A_0\bar A_1})^2=-\delta t^2 \Tr_{\bar A_0} [\Tr_{A_0}(\rho_0|I\rangle \langle J|H)\Tr_{A_0}(\rho_0|J\rangle \langle I|H)]\nn \\
&&\phantom{tr(T^{(1)}_{A_0\bar A_1})^2=}
+2\delta t^2 \Tr_{\bar A_0} [\Tr_{A_0}(\rho_0|I\rangle \langle J|H)\Tr_{A_0}(H\rho_0|J\rangle \langle I|)]-\delta t^2 \Tr(H\rho_0)^2\nn \\
&&2tr(T^{(0)}_{A_0\bar A_1}T^{(2)}_{A_0\bar A_1})=-\delta t^2 \Tr_{\bar A_0} [\Tr_{A_0}(\rho_0|I\rangle \langle J|H^2)\Tr_{ A_0}(\rho_0|J\rangle \langle I|)]-\delta t^2 \Tr (H^2 \rho_0^2)\nn \\
&&\phantom{2tr(T^{(0)}_{A_0\bar A_1}T^{(2)}_{A_0\bar A_1})=}+2\delta t^2 \Tr_{\bar A_0} [\Tr_{A_0}(H\rho_0|I\rangle \langle J|H)\Tr_{ A_0}(\rho_0|J\rangle \langle I|)].
\eea

For the moment $tr(T_{A_0A_1})^2$, the $O(\delta t)$ result is quite simple. For the special case where $\rho_0$ is a product, $\rho_0=\rho_{A_0}\otimes \rho_{\bar A_0}$, the result vanishes. In the next section, we will see that the form of the interaction is also sensitive to the leading-order result of the moment. It is also interesting that $tr(T_{A_0A_1})^2$ vanishes at leading order for the product density matrix. This follows from the fact that for $\rho_0=\rho_{A_0}\otimes \rho_{\bar A_0}$
\bea
&&\Tr_{\bar A_0}(\Tr_{A_0}(\rho_0|I\rangle \langle J|H)\Tr_{A_0}(\rho_0|J\rangle \langle I|))\nn \\
&&=\Tr_{\bar A_0}(\rho_{\bar A_0}\Tr_{A_0}(\rho_{A_0}|I\rangle \langle J|H)\rho_{\bar A_0} \langle I| \rho_{A_0}|J\rangle)\nn \\
&&=\Tr_{\bar A_0}(\rho_{\bar A_0}\Tr_{A_0}(\rho_{A_0}^2 H)\rho_{\bar A_0}) \nn \\
&&=\Tr(\rho_{A_0}^2\otimes \rho_{\bar A_0}^2 H).
\eea
Using the above formula, one can see that the order $O(\delta t)$ result of $tr(T_{A_0\bar A_1})^2$ vanishes. 

In Appendix~\ref{Appendix_moments_reduced}, we also study $tr(T_{A_0A_1}T_{A_0A_1}^\dagger)$ and $tr(T_{A_0\bar A_1}T_{A_0\bar A_1}^\dagger)$ up to $O(\delta t)$. In general, the second moments of the reduced spacetime density matrices exhibit similar behavior in the short-time limit. They appear to depend on the choice of basis, the Hamiltonian, and the initial density matrix. The leading-order results at $O(\delta t)$ can be straightforwardly generalized to $n$-th moments. However, for arbitrary $n$ the next-to-leading-order results are more complicated to evaluate using perturbation methods.

\subsection{Sensitivity to couplings between subsystems}
Consider two subsystems $A_0$ and $\bar A_0$ with the Hamiltonian (\ref{Hamiltonian}). In the limit $\lambda\to 0$, there is no interaction between the two subsystems. We will show below that in this limit the moments of $T_{A_0A_1}$ and $T_{A_0\bar A_1}$ give trivial results that are independent of time $t$. In this limit, the evolution operator factorizes as $U = U_{A_0} U_{\bar A_0}$, which corresponds to local unitary operations on the subsystems $A_0$ and $\bar A_0$. It is a well-known result that, for spacelike-separated subsystems, the entanglement entropy is invariant under local unitary operations. Here, however, we consider arbitrary spacetime subsystems, and the situation is different. In some sense, this can be seen as a generalization to arbitrary subsystems.

\subsubsection{Trivial results without coupling}\label{Section_trivial}
In the limit $\lambda\to 0$, we have
\bea\label{TA0A1_first}
&&T_{A_0A_1}=\langle L|\langle \bar K|U_{A_0} U_{\bar A_0} \rho_0|I\rangle |\bar I\rangle \langle J| U_{A_0}^\dagger|K\rangle  \langle \bar I| U_{\bar A_0}^\dagger |\bar K\rangle |J\rangle \langle I|\otimes |L\rangle \langle K|\nn \\
&&\phantom{T_{A_0A_1}}=\langle L|\langle \bar I|U_{A_0}\rho_0|I\rangle|\bar I\rangle \langle J|U_{A_0}^\dagger|K\rangle |J\rangle \langle I|\otimes |L\rangle \langle K|\nn \\
&&\phantom{T_{A_0A_1}}=\langle L|U_{A_0}\rho_{A_0}|I\rangle\langle J|U_{A_0}^\dagger|K\rangle |J\rangle \langle I|\otimes |L\rangle \langle K|
\eea
where we use the complete relation $|\bar I\rangle \langle \bar I|=\bar 1$ (the identity in the Hilbert space $H_{\bar A_0}$). It is more convinent to write the above expression as
\bea\label{reduced_A0system}
T_{A_0A_1}=U_{A_0}^\dagger |K\rangle \langle L|U_{A_0}\rho_{A_0}\otimes |L\rangle \langle K|.
\eea
Compared with (\ref{Two_Form1}), $T_{A_0A_1}$ takes the same form as the spacetime density matrix $T_{C_0C_1}$. This is expected since the two systems are decoupled and thus evolve independently. $T_{A_0A_1}$ is just the spacetime density matrix for one of the subsystems, with the initial state $\rho_{A_0}$ and evolution operator $U_{A_0}$. Similarly, we can define and compute the reduced operator $T_{\bar A_0\bar A_1}$, which is given by
\bea\label{reduced_barA0system}
T_{\bar A_0\bar A_1}=U_{\bar A_0}^\dagger |\bar K\rangle \langle \bar L|U_{\bar A_0}\rho_{\bar A_0}\otimes |\bar L\rangle \langle \bar K|.
\eea
However, it should be noted that 
\bea
T_{C_0C_1} \ne T_{A_0 A_1} \otimes T_{\bar A_0 \bar A_1},
\eea
unless $\rho_0 = \rho_{A_0} \otimes \rho_{\bar A_0}$. This indicates that the reduced transition operator $T_{A_0A_1}$ does reflect the entanglement between different spacetime subregions.

We can also consider $T_{A_0\bar A_1}$. We have
\bea\label{TA0barA1_first}
&&T_{A_0\bar A_1}=\langle K |\langle \bar L|U_{A_0}U_{\bar A_0}\rho_0|I\rangle|\bar I\rangle \langle J| \langle \bar I|U_{A_0}^\dagger U_{\bar A_0}^\dagger|K\rangle |\bar K\rangle |J\rangle\langle I| \otimes |\bar L\rangle \langle \bar K|\nn \\
&&\phantom{T_{A_0\bar A_1}}=\langle J| \langle \bar L| U_{\bar A_0}\rho_0 U_{\bar A_0}^\dagger |\bar K\rangle |I\rangle  |J\rangle \langle I|\otimes |\bar L\rangle \langle \bar K|,
\eea
which corresponds to $U_{\bar A_0}\rho_0 U_{\bar A_0}^\dagger$ in the basis $|I\rangle \otimes |\bar I\rangle$. This result indicates that the operator $T_{A_0\bar A_1}$ refers to the density matrix of the subsystems at $A_0$ and $A_1$.

It is clear that $tr(T_{A_0\bar A_1})^n$ is given by $\Tr \rho_{0}^n$, which depends only on the moments of the initial density matrix. Hence, the result is independent of time. In this case, the entanglement entropy for the subsystems $A_0$ and $A_1$ is trivial due to the absence of interaction between the two subsystems. Similarly, $tr(T_{A_0A_1})^n$ is also independent of time, and the result, which can be calculated using (\ref{n_moment}), depends only on the initial state $\rho_0$.

Therefore, only when the interaction between the two subsystems is turned on do the moments of $T_{A_0A_1}$ and $T_{A_0\bar A_1}$ give non-trivial results. This indicates that the entropy-related quantities of the reduced transition operators can be used as a probe to study the interactions between subsystems.

\subsubsection{Consistent check}

In Section~\ref{section_perturbation_short}, we calculate the moments of $T_{A_0A_1}$ and $T_{A_0\bar A_1}$ in the short-time limit for an arbitrary Hamiltonian $H$. If we set the coupling $\lambda=0$, it is expected that the first-order perturbation results (\ref{perturbation_short_TA0A1}) and (\ref{perturbation_short_TA0barA1}) in the short-time limit would vanish. In the limit $\lambda \to 0$, the Hamiltonian reduces to $H=H_{A_0}+H_{\bar A_0}$.
For $T_{A_0A_1}$, the first-order result is
\bea
&&2tr(T^{(0)}_{A_0A_1}T^{(1)}_{A_0A_1})\nn \\
&&=-2i\delta t \Tr\left[(\rho_0-\rho_{A_0}\otimes \rho_{\bar A_0})H\right]\nn\\
&&=-2i\delta t \Tr\left[\rho_0-\rho_{A_0}\otimes \rho_{\bar A_0})H_{A_0}\right]-2i\delta t \Tr\left[\rho_0-\rho_{A_0}\otimes \rho_{\bar A_0})H_{\bar A_0}\right]=0.\nn
\eea
For $T_{A_0\bar A_1}$, the first order result is (\ref{perturbation_short_TA0barA1}). Let us consider the term
\bea
&&\Tr_{\bar A_0}[\Tr_{A_0}(\rho_0|I\rangle \langle J|H)\Tr_{A_0}(\rho_0|J\rangle \langle I|)]\nn \\
&&=\Tr_{\bar A_0}\left[\left(\Tr_{A_0}(\rho_0|I\rangle \langle J|H_{A_0})+\Tr_{A_0}(\rho_0|I\rangle \langle J|H_{\bar A_0})\right) \langle I|\rho_0|J\rangle \right]\nn \\
&&=\Tr_{\bar A_0}\left[ \langle J|H_{A_0}\rho_0|I\rangle \langle I|\rho_0|J\rangle+ \langle J| \rho_0|I\rangle H_{\bar A_0} \langle I|\rho_0|J\rangle \right]\nn \\
&&=\Tr (H_{A_0}\rho_0^2)+\Tr (H_{\bar A_0}\rho_0^2)\nn\\
&&=\Tr(H\rho_0^2).
\eea
Thus, the first order result in (\ref{perturbation_short_TA0barA1}) is vanishing. It is also straightforward to consider the second order of $O(\delta t^2)$, and check the results are vanishing in the limit $\lambda\to 0$.

\subsubsection{Perturbative calculation}\label{Section_perturbation_method}
If the dimensionless coupling constant $\lambda \ll 1$, one can evaluate the moments using a perturbative method. The idea is to expand the evolution operator $U = e^{-i H t}$ in a series in $\lambda$. There exists a standard method to perform a similar expansion in perturbative QFTs.

Now let us consider the spacetime density matrix $T_{C_0C_1}$ and its reduced operators, which involve the evolution operator $U(t_1,t_0) = e^{-i H (t_1 - t_0)}$. In this section, we will set $t_0 = 0$ and $t_1 = t$. With the Hamiltonian (\ref{Hamiltonian}), the evolution operator $U(t_1,t_0)$ can be rewritten as
\bea
U(t,0)=e^{-i H_0 t} U_{I}(t),
\eea
where $U_I$ satisfies the following equation
\bea\label{dyson}
i\frac{d}{dt}U_I(t)=\lambda V_I(t) U_I(t),
\eea
with $V_I:=e^{iH_0 t}Ve^{-iH_0 t}$, $H_0$ is the free Hamiltonian defined by (\ref{free_Hamiltonian}).
It is obvious that $U_I$ satisfies the initial condition $U_I(0) = I$. 
The above equation indicates that $U_I$ is simply the evolution operator in the interaction picture, which is generally used to handle the interaction terms in a quantum system. Eq.~(\ref{dyson}) has the standard Dyson series solution:
\bea
U_I(t) = T e^{-i\lambda \int_0^t dt' \, V_I(t')},
\eea
where $T$ denotes the time-ordering operator. In the special case where $[H_0, V] = 0$, $V_I$ is time-independent, and thus $U_I = e^{-i\lambda V t}$. The full evolution operator can then be written as $U = e^{-i H_0 t} e^{-i \lambda V t}$, which can also be obtained using the Baker–Campbell–Hausdorff (BCH) formula.

In the perturbative calculation, one can use the Dyson expansion:
\bea\label{Dyson_expansion}
U_I(t)=1-i\lambda \int_0^{t}dt_1  V_I(t_1)+(-i\lambda)^2 \int_0^tdt_1 \int_0^{t_1}dt_2 V_{I}(t_1)V_I(t_2)+\cdots\; .
\eea
In the following, we will only consider the first-order perturbation. 

Our motivation is to study the relation between the moments of $T_{A_0A_1}$ and $T_{A_0\bar A_1}$ and the interaction. Up to $O(\lambda)$, the operator $T_{A_0A_1}$ is given by
\bea
T_{A_0A_1}=T^{[0]}_{A_0A_1}+\lambda T^{[1]}_{A_0A_1}+\cdots,
\eea
where $T^{[0]}_{A_0A_1}$ is given by (\ref{TA0A1_first}) and
\bea
&&T^{[1]}_{A_0A_1}=\langle L|\langle \bar I|U_{A_0}U_I^{[1]}\rho_0|\bar I\rangle |I\rangle \langle J|U_{A_0}^\dagger|K\rangle |J\rangle \langle I|\otimes |L\rangle \langle K|\nn \\
&&\phantom{T^{[1]}_{A_0A_1}=}+\langle L|\langle \bar K| U_{A_0}U_{\bar A_0}\rho_0|I\rangle |\bar I\rangle \langle J| \langle \bar I| (U_{I}^{[1]})^\dagger U_{A_0}^\dagger U_{\bar A_0}^\dagger |K\rangle |\bar K\rangle |J\rangle \langle I|\otimes |L\rangle \langle K|,\nn
\eea
where $U_{A_0}=e^{-iH_{A_0}t}$ and $U_{\bar A_0}=e^{-iH_{\bar A_0}t}$, $U_I^{[1]}=-i\int_0^tdt_1V_I(t_1)$.

While the reduced operator $T_{A_0\bar A_1}$ is given by
\bea
T_{A_0\bar A_1}=T^{[0]}_{A_0\bar A_1}+\lambda T^{[1]}_{A_0\bar A_1}+\cdots,
\eea
where $T^{[0]}_{A_0\bar A_1}$ is given by (\ref{TA0barA1_first}) and
\bea
&&T^{[1]}_{A_0\bar A_1}=\langle J|\langle \bar L|U_{\bar A_0}U_I^{[1]}\rho_0U_{\bar A_0}^\dagger|I\rangle |\bar K\rangle |J\rangle \langle I|\otimes |\bar L\rangle \langle \bar K|\nn \\
&&\phantom{T^{[1]}_{A_0\bar A_1}=}+\langle K|\langle \bar L| U_{A_0}U_{\bar A_0}\rho_0|I\rangle |\bar I\rangle \langle J|\langle \bar I| (U_I^{[1]})^\dagger U_{A_0}^\dagger U_{\bar A_0}^\dagger |K\rangle |\bar K\rangle |J\rangle \langle I|\otimes |\bar L\rangle \langle \bar K|.\nn \\
~
\eea
Now we are ready to calculate the moments. For $T^{[0]}_{A_0A_1}$ we have
\bea\label{perturbation_lambda_TA0A1}
tr(T_{A_0A_1})^2=tr (T^{[0]}_{A_0A_1})^2+2\lambda tr(T^{[0]}_{A_0A_1}T^{[1]}_{A_0A_1})+\cdots,
\eea
with 
\bea
&&tr (T^{[0]}_{A_0A_1})^2= (\Tr\rho_{A_0})^2,\nn \\
&&2\lambda tr(T^{[0]}_{A_0A_1}T^{[1]}_{A_0A_1})= 2\lambda \left[\Tr \rho_0 U_I^{[1]}+\Tr (\rho_{A_0}\otimes \rho_{\bar A_0} (U_I^{[1]})^\dagger)\right]\nn \\
&&\phantom{2\lambda tr(T^{[0]}_{A_0A_1}T^{[1]}_{A_0A_1})}=-2i\lambda \int_0^tdt_1 \Tr\left[(\rho_0-\rho_{A_0}\otimes \rho_{\bar A_0})V_{I}(t_1)\right].
\eea
For $T^{[0]}_{A_0\bar A_1}$ we have
\bea\label{perturbation_lambda_TA0barA1}
tr(T_{A_0\bar A_1})^2=tr (T^{[0]}_{A_0\bar A_1})^2+2\lambda tr(T^{[0]}_{A_0\bar A_1}T^{[1]}_{A_0\bar A_1})+\cdots\; ,
\eea
with 
\bea
&&tr (T^{[0]}_{A_0\bar A_1})^2= \Tr\rho_0^2,\nn \\
&&2\lambda tr(T^{[0]}_{A_0\bar A_1}T^{[1]}_{A_0\bar A_1})\nn \\
&&=2\lambda\left[ \Tr (U_I^{[1]}\rho_0^2)+\Tr_{\bar A_0}\left( \Tr_{A_0}(\rho_0 |I\rangle \langle K|U_{A_0}) \Tr_{A_0} (U_{I}^{[1]})^\dagger U_{A_0}^\dagger |K\rangle \langle I|\rho_0 \right)\right]\nn \\
&&=-2i\lambda\int_0^tdt_1\left[ \Tr (V_I(t_1)\rho_0^2)-\Tr_{\bar A_0}\left( \Tr_{A_0}(\rho_0 |I\rangle \langle K|U_{A_0}) \Tr_{A_0}(V_I(t_1) U_{A_0}^\dagger |K\rangle \langle I|\rho_0) \right)\right].\nn\\
~ 
\eea
In principle, one could calculate the $n$-th moment of $T_{A_0 A_1}$ and $T_{A_0\bar A_1}$ for arbitrary integers $n$ at the leading order in $O(\lambda)$. It is also straightforward to compute higher orders in $\lambda$ using the Dyson expansion (\ref{Dyson_expansion}). However, the operator perturbation method becomes increasingly complicated at higher orders. In Section~\ref{Section_discussion}, we will briefly discuss the procedure for evaluating the moments using a diagrammatic method.

The result of $tr(T_{A_0A_1})^2$ (\ref{perturbation_lambda_TA0A1}) at leading order in $\lambda$ is simple and depends on the difference between the expectation value of the interaction in $\rho_0$ and that in the product form $\rho_{A_0}\otimes \rho_{\bar A_0}$. This result is non-vanishing unless $\rho_0$ is a product state, i.e., $\rho_0= \rho_{A_0}\otimes \rho_{\bar A_0}$. Therefore, $tr(T_{A_0A_1})^2$ serves as a good probe to detect interactions between the two subsystems.

The result for $tr(T_{A_0\bar A_1})^2$ (\ref{perturbation_lambda_TA0barA1}) is more complicated for a general initial density matrix $\rho_0$. However, if the initial density matrix is a product state, i.e., $\rho_0=\rho_{A_0}\otimes \rho_{\bar A_0}$, the term of order $O(\lambda)$ in (\ref{perturbation_lambda_TA0barA1}) vanishes:
\bea
2\lambda tr(T^{[0]}_{A_0\bar A_1}T^{[1]}_{A_0\bar A_1})=-2i\lambda \int_0^tdt_1\left[ \Tr (V_I(t_1)\rho_0^2)-\Tr (V_I(t_1) \rho_{A_0}^2\otimes \rho_{\bar A_0}^2)  \right]=0.
\eea
Thus the result is also vanishing for the initial product density matrix.  

For both $T_{A_0A_1}$ and $T_{A_0\bar A_1}$, the leading-order result of the 2-nd moment vanishes for the product density matrix $\rho_{A_0}\otimes \rho_{\bar A_0}$. However, higher-order results are generally non-vanishing. The interaction generates entanglement between the two systems. In this sense, the moments of $T_{A_0A_1}$ and $T_{A_0\bar A_1}$ may serve as useful quantities to detect the generation of entanglement due to the interaction between the two systems.

\subsection{Remarks on entanglement for causally connected subregions in QFTs}

An important application of the (reduced) spacetime density matrix $T_{C_0C_1\cdots C_i\cdots C_{N-1}}$ is to define entanglement for causally connected subregions in QFTs. This was one of the motivations in \cite{Milekhin:2025ycm}, where they briefly discuss the two-subregion case. In \cite{Guo:2022jzs}, we discuss in more detail the Schwinger-Keldysh formalism of the transition operator and the replica method to evaluate the entanglement entropy for causally connected subregions in more general cases. When considering entanglement among subregions on multiple Cauchy surfaces, it is necessary to define $T_{C_0C_1\cdots C_i\cdots C_{N-1}}$ as in (\ref{definition_T0N}).

The definitions of the transition operator and its reduced form used in this paper can be applied in QFTs, and many of the results can also be extended to QFTs. Some results, however, depend on the choice of basis for subsystems $A_0$ and $\bar A_0$, such as the perturbative results in Section~\ref{section_perturbation_short}. In QFTs, these can only be considered formal expressions, since it is generally difficult or impossible to obtain a basis for subregions. Moreover, the algebraic structure in QFTs is quite different from that of finite-dimensional quantum systems. Even with certain regularizations, while some information about projectors for reduced density matrices in subregions can be obtained \cite{Guo:2024jyr}, it remains unclear how to directly compute quantities like the moments of $T_{A_0A_1}$ using projectors.

However, in QFTs one can use the Schwinger-Keldysh path integral and the real-time replica method. As a result, the moments of $T_{A_0A_1}$ or more general $T_{A_0A_1\cdots A_i\cdots A_{N-1}}$ can be evaluated using the twist operator formalism. In general, this reduces to computing correlation functions involving twist operators in Lorentzian spacetime. In QFTs, standard methods exist to calculate these correlation functions via analytic continuation from Euclidean results. Therefore, the method to evaluate entropy-related quantities for reduced transition operators in QFTs is quite different from the approach used in this paper. One can refer to \cite{Guo:2022jzs} for more details on the calculations in QFTs.

\subsection{Examples}\label{Section_example}
It is useful to illustrate the evaluation of the moments of the reduced spacetime density matrix using some examples. Here, we consider two simple cases: a two-qubit system and the thermal field double (TFD) state.

\subsubsection{Two qubits system}
\begin{figure}[htbp]
  \centering
  \includegraphics[width=0.5\textwidth]{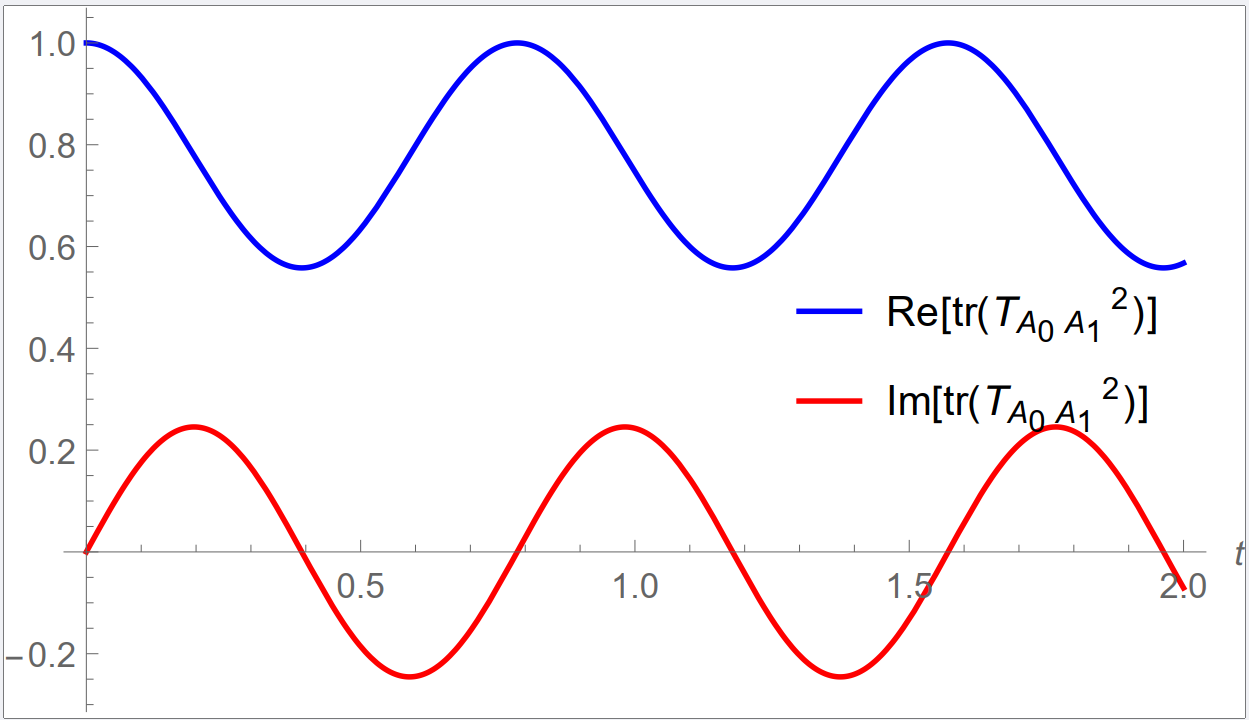}
  \caption{Plot of $trT_{A_0A_1}^2$. We set $J=1$ and $\theta=\frac{\pi}{6}$. 
}\label{Fig_two_qubit_TA0A1}
\end{figure}

\begin{figure}[htbp]
    \centering
    \subfigure{
        \includegraphics[width=0.45\textwidth]{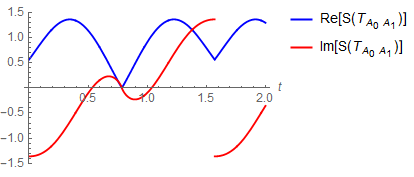}}
    \subfigure{
        \includegraphics[width=0.45\textwidth]{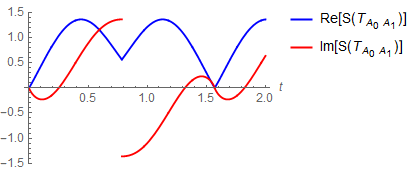}}
        \caption{ Plot of the entanglement entropy for $T_{A_0A_1}$ and $T_{A_0B_1}$ in the two qubits model. We set $J=1$ and $\theta=\frac{\pi}{6}$}\label{Fig_two_qubit_EE}
\end{figure}

Let us consider a two-qubit system consisting of Alice ($A$) and Bob ($B$) with the Heisenberg coupling:
\bea
H=J(\sigma_x\otimes \sigma_x+\sigma_y\otimes \sigma_y+\sigma_z\otimes \sigma_z),
\eea
where $J$ is real constants, $\sigma_i$ ($i=x,y,z$) are Pauli matrices. The basis of the systems are $\{|00\rangle,|01\rangle,|10\rangle,|11\rangle\}$. Let us consider the initial state:
\bea
|\psi_0\rangle=\cos\theta |01\rangle+\sin\theta |10\rangle.
\eea
We consider the systems at times $t=0$ and $t$. Alice and Bob at time $t=0$ and $t$ are denoted by $A_0, B_0$ and $A_1, B_1$, respectively.  
It is straightforward to obtain $T_{C_0C_1}$ using (\ref{Transition_two}). Then, we can calculate the reduced operators $T_{A_0A_1}$ and $T_{A_0B_1}$ using the formula (\ref{reduced_two_party}). The results are complicated and are shown in Appendix~\ref{Appendix_twoqubits}.  
We can also evaluate the second moments; in this example, we have\footnote{For a general initial state, we do not expect $tr(T_{A_0A_1})^2 = tr(T_{A_0B_1})^2$. This example is specific.}
\bea
&&tr(T_{A_0A_1})^2=tr(T_{A_0B_1})^2\nn \\
&&=\frac{1}{8} \Big(2 \sin (2 \theta )+i (4 \sin (2 \theta )-1) \sin (8 J t)+\cos (4 \theta ) \left(1-e^{-8 i J t}\right)\nn \\
&&\phantom{=}+(3-2 \sin (2 \theta )) \cos (8 J t)+5\Big).\nn
\eea
We can expand the above results in the short-time limit and find consistency with the results in (\ref{perturbation_short_TA0A1}) and (\ref{perturbation_short_TA0barA1}).  
It is also notable that the results depend non-trivially on time, even for the product states with $\theta = 0, \frac{\pi}{2}$.  
An illustration of the second moment is given in Fig.~\ref{Fig_two_qubit_TA0A1}.
We can also obtain 
\bea
&&tr(T_{A_0A_1}T_{A_0A_1}^\dagger)=\frac{1}{8} \left(\cos (8 J t)+9+\cos (4 J t) \left(4 \cos (4 \theta ) \cos ^2(2 J t)+2\right)\right),\nn \\
&&tr(T_{A_0B_1}T_{A_0B_1}^\dagger)= \frac{1}{8} \left(\cos (8 J t)+9-2 \cos (4 J t) \left(2 \cos (4 \theta ) \sin ^2(2 J t)+1\right)\right).
\eea
In this simple model we can also obtain the entanglement entropy for $T_{A_0A_1}$ and $T_{A_0B_1}$, which is defined as $$S(T_{A_0A_1})=-tr\log T_{A_0A_1}\log T_{A_0A_1},\quad S(T_{A_0B_1})=-tr\log T_{A_0B_1}\log T_{A_0B_1}.$$
We plot the entanglement entropy for the two reduced transition operators in Fig.~\ref{Fig_two_qubit_EE}.  
A notable feature is that the imaginary part of the entanglement entropy exhibits a sharp sign inversion at certain times, as can be seen from the plot.

\subsubsection{Thermal field double state}\label{Section_TFD}

The thermal field double (TFD) state provides a purification of the canonical thermal state $e^{-\beta H}$ by introducing an auxiliary copy of the system. Let $H_L$ and $H_R$ denote the Hamiltonians of the two copies. In the energy eigenbasis ${|E_i\rangle_L}$ and ${|E_i\rangle_R}$, the TFD state is defined as
\bea
|\Psi\rangle = \frac{1}{\sqrt{Z(\beta)}} \sum_i e^{-\frac{\beta}{2}E_i}  |E_i\rangle_L \otimes |E_i\rangle_R ,
\eea
where $Z(\beta)=\sum_i e^{-\beta E_i}$ is the partition function. This state is entangled between the two copies. Tracing out one subsystem from the density matrix $|\Psi\rangle\langle\Psi|$ yields the thermal state of the other subsystem.

For the total Hamiltonian $H=H_L-H_R$, one finds $H|\Psi\rangle=0$, implying that the TFD state is time-independent. Alternatively, if we consider $H'=H_L+H_R$, the state evolves nontrivially in time. The TFD state is also believed to be dual to the eternal black hole in AdS spacetime \cite{Maldacena:2001kr}, and has been widely used to probe black hole properties \cite{Maldacena:2013xja}–\cite{Brown:2015bva}. In this duality, the entanglement entropy of $L$ or $R$ in the TFD state corresponds to the black hole entropy.

\begin{figure}[htbp]
  \centering
  \includegraphics[width=0.7\textwidth]{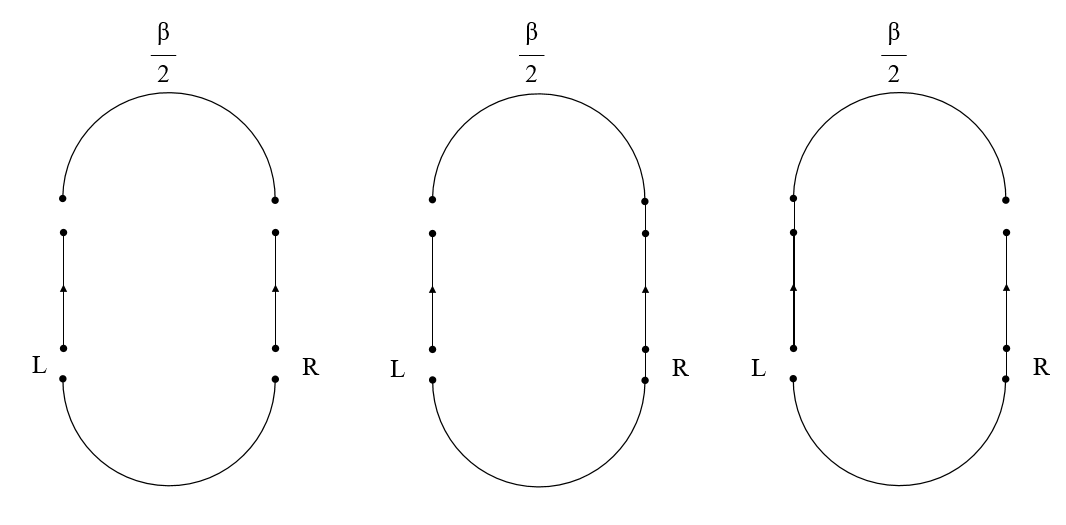}
  \caption{Path representation of the operators $\mathcal{F}_{C_0C_1,s}^\dagger$ (left), $\mathcal{F}_{L_0L_1,s}^\dagger$ (middle) and $\mathcal{F}_{L_0R_1,s}^\dagger$ (right).
}\label{Fig_SK_TFD}
\end{figure}

The two systems $L$ and $R$ provide a natural example to construct reduced spacetime density matrix and study entropy-related quantities. Here, we consider the Hamiltonian $H'$, so that both systems evolve in the same time direction. By considering two Cauchy surfaces at $t=0$ and $t$, one can construct the operator $T_{C_0C_1}$, as well as the reduced transition operators $T_{L_0L_1}$ and $T_{L_0R_1}$. However, according to the results in Section~\ref{Section_trivial}, the moments of $T_{L_0L_1}$ and $T_{L_0R_1}$ are trivial, since the Hamiltonian $H'$ contains no interaction term between $L$ and $R$. More precisely, we have
\bea
tr(T_{L_0L_1})^n=\begin{cases}
\frac{Z(n\beta)}{Z(\beta)^n}, & n \in \text{odd},  \\
\frac{Z(n\beta/2)^2}{Z(\beta)^n},  & n \in  \text{even},
\end{cases}
\eea
and
\bea
tr(T_{L_0R_1})^n=1.
\eea
If one wishes the results to exhibit nontrivial time dependence, there are two possible approaches. One approach is to use the generalized transition operator introduced in Section~\ref{Section_generalization}. Here, we consider the operator $\mathcal{F}_{C_0C_1,s}^\dagger$ with the state choice $|\psi_0\rangle = |\psi_1\rangle = |\Psi(\beta)\rangle$. The explicit form of this operator is given in Eq.~(\ref{Generalization_F}). For the TFD state under consideration, one can also define the corresponding reduced operator
\bea
&&\mathcal{F}_{L_0L_1,s}^\dagger=tr_{R_0R_1}\mathcal{F}_{C_0C_1,s}^\dagger,\nn \\
&&\mathcal{F}_{L_0R_1,s}^\dagger=tr_{R_0L_1}\mathcal{F}_{C_0C_1,s}^\dagger.
\eea
We can expression these operators by path integral as shown in Fig.\ref{Fig_SK_TFD}.

 With this expression it is straightforward to evlauate the moments 
\bea
tr(\mathcal{F}_{L_0L_1,s}^\dagger)^n,\quad tr(\mathcal{F}_{L_0R_1,s}^\dagger)^n.
\eea
The $2$-nd moment is shown in Fig.\ref{Fig_SK_2moment}. In Appendix.\ref{Appendix_TFD} we show more examples of higher-order moments.
\begin{figure}[htbp]
    \centering
    \subfigure{
        \includegraphics[width=0.6\textwidth]{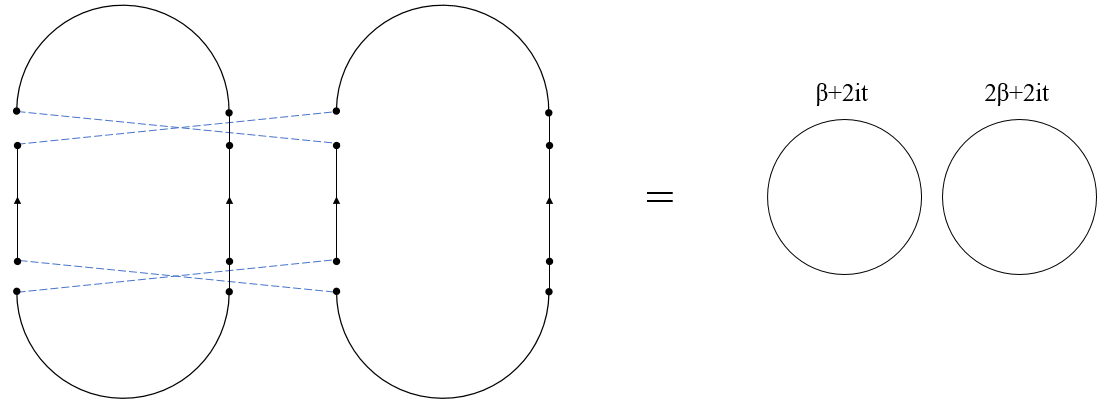}}
    \subfigure{
        \includegraphics[width=0.6\textwidth]{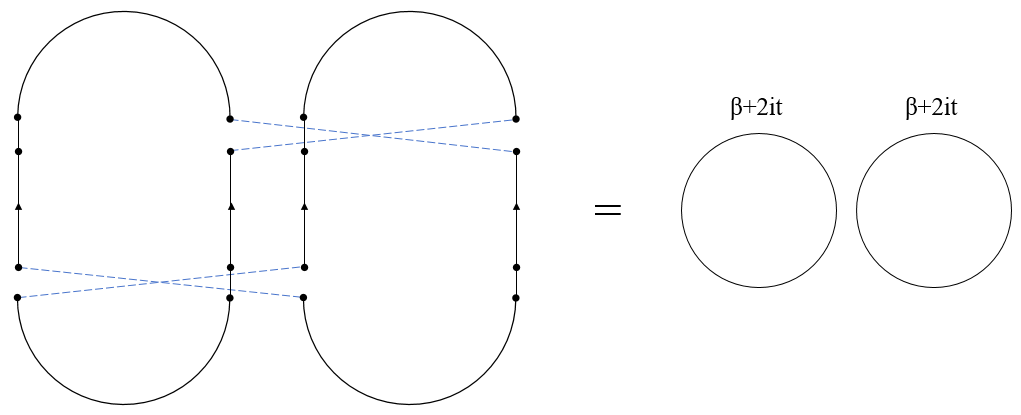}}
        \caption{Path integral representation of the 2-nd moment $tr(\mathcal{F}_{L_0L_1,s}^\dagger)^2,\quad tr(\mathcal{F}_{L_0R_1,s}^\dagger)^2$.}\label{Fig_SK_2moment}
\end{figure}

As a result we find
\bea
&&tr(\mathcal{F}_{L_0L_1,s}^\dagger)^n=\begin{cases}
Z(n(\beta+2it)), & n \in \text{odd},  \\
Z(\frac{n}{2}(\beta+2it))^2,  & n \in  \text{even}.
\end{cases}\nn\\
&&tr(\mathcal{F}_{L_0R_1,s}^\dagger)^n=Z(\beta+2it)^n.
\eea

Another way to introduce time dependence is by including an interaction between the $L$ and $R$ systems. Motivated by the study of traversable wormholes in eternal black holes, the authors of \cite{Gao:2016bin} introduce an interaction that couples the two CFTs living on the boundaries of the eternal black hole. We focus on the AdS$_3$ case, where the CFTs are two-dimensional and living on a circle. Here, we consider a constant interaction term in the Hamiltonian, given by
\bea
V=-\int d\phi h(\phi) \mathcal{O}_R(0,\phi) \mathcal{O}_L(0,\phi),
\eea
where $h(\phi)$ is a smooth function controlling the interaction strength, and $\mathcal{O}_L$ and $\mathcal{O}_R$ are primary operators living on the left and right boundary CFTs, respectively. We consider the Hamiltonian to be $H = H_R - H_L$. By this definition, we then have
\bea
V_I(t)=e^{-iH t}V e^{iHt}=-\int d\phi h(\phi) \mathcal{O}_R(t,\phi) \mathcal{O}_L(-t,\phi).
\eea
Here, we focus solely on $T_{A_0A_1}$. Assuming the interaction is weak, the moments of $T_{A_0A_1}$ can be evaluated using formula (\ref{perturbation_lambda_TA0A1}), yielding
\bea
&&tr(T_{A_0A_1}^2)=1+2i\int_0^tdt_1 \int d\phi h(\phi) \Tr\left[(\rho_0-\rho_{A_0}\otimes \rho_{\bar A_0})\mathcal{O}_R(t_1,\phi) \mathcal{O}_L(-t_1,\phi)\right]+O(\lambda^2),\nn\\
&&\phantom{tr(T_{A_0A_1}^2)}=1+2i\int_0^tdt_1 \int d\phi h(\phi) \langle \Psi(\beta)|\mathcal{O}_R(t_1,\phi) \mathcal{O}_L(-t_1,\phi)|\Psi(\beta)\rangle+O(\lambda^2),
\eea
where the one-point function of the primary operator vanishes. The result depends only on the two-point functions of the left and right operators, which can be evaluated directly using the KMS condition for the TFD state. A more detailed study on this topic will be presented elsewhere \cite{Guo:future}.

\section{Conclusion and Outlook}\label{Section_discussion}

In this paper, we investigate the concept of the spacetime density matrix. Its definition is fundamental, as the construction can be applied to arbitrary quantum systems. Our formalism is developed in a general setting, motivated by the study of the properties of the moments of the spacetime density matrix and its reduced forms. These moments, or related entropy-like quantities, have been shown to provide valuable probes of the dynamics of a system and its interactions with other systems.

In QFTs, the focus is typically on subregions of spacetime, where it is not straightforward to decompose the Hamiltonian into ``subsystem Hamiltonians''. The TFD state discussed in Section~\ref{Section_TFD} offers a natural example within our framework. When the Hamiltonian is taken as $H = H_R \pm H_L$, the two CFTs are non-interacting, and the reduced spacetime density matrix becomes trivial, in accordance with the result derived in Section~\ref{Section_trivial}. However, introducing interactions between the two CFTs can make the situation nontrivial, and such interactions are directly related to traversable wormholes, as discussed in \cite{Gao:2016bin}. This model thus provides a particularly interesting setting in which to study the moments or entropy-like quantities of reduced spacetime density matrices such as $T_{L_0R_0}$ and $T_{L_0R_1}$. We plan to pursue further investigations of this model in future work, including possible holographic duals of the entropy-related quantities.

The framework presented here may also find broader applications in the study of open quantum systems \cite{Breuer}, where the system interacts with an environment through an interaction term $V$ in the Hamiltonian. In earlier studies, this setup naturally led to a Lindblad-type equation of motion for the system’s density matrix. We show here that the spacetime density matrix satisfies a Liouville–von Neumann–type equation of motion. For the reduced spacetime density matrix of the system, it would be interesting to determine what form of evolution equation it obeys.

The perturbative method developed in Section~\ref{Section_perturbation_method} can be applied to calculate the moments of the reduced spacetime density matrix when the coupling between two systems is weak. In this paper we have only carried out a leading-order calculation. In principle, the computation can be systematically extended to higher orders, as in perturbative QFTs where standard methods exist. The main challenge here is how to systematize the calculation.

In QFT, the computation of scattering amplitudes relies fundamentally on Feynman diagrams. Similarly, the quantities we are concerned with here can also be computed using a diagrammatic representation. For the TFD state example, when the interaction is turned off, we indeed employ diagrams to compute the moments of $T_{L_0L_1}$ and $T_{L_0R_1}$, although the results are trivially independent of time. Once interactions are introduced, however, it becomes possible to modify the diagrams and develop a set of rules analogous to Feynman rules. This provides a promising approach to computing higher-order corrections to the moments.

In this paper we have mainly focused on the general formalism of the spacetime density matrix, but its physical interpretation remains unclear. A particularly important direction is to understand the physical meaning of the new quantities defined by the spacetime density matrix. It would be especially interesting to apply these quantities to scenarios involving interactions, such as scattering processes or decoherence due to coupling with an environment \cite{Zurek:1981xq,Zurek:1982ii}.

~\\

{\bf Acknowledgements}
I would like to thank Bin Chen, Peng Cheng, Yun-Gui Gong, Song He, Yu-Xiao Liu, Jian-Xin Lu, Hong Lu,  Rong-Xin Miao, Carlos Nunez, Yi Pang, Dibakar Roychowdhury, Fu-Wen Shu,Jie-Ci Wang, Jie-qiang Wu, Run-qiu Yang, Hong-bao Zhang, Jiaju Zhang, Yang Zhou and  Yu-Xuan Zhang for useful discussions. I am especially grateful to Li-Yuan Gao for assistance in preparing the figures for this paper.
I am supposed by the National Natural Science Foundation of China under Grant No.12005070 and the Hubei Provincial Natural Science Foundation of China under Grant No.2025AFB557.

\appendix

\section{Another Generalization}\label{Section_another}
\begin{figure}[htbp]
  \centering
  \includegraphics[width=0.23\textwidth]{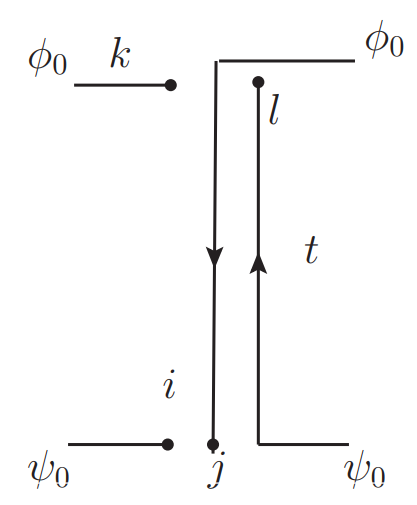}
  \caption{Path integral representation of the operator $\mathcal{D}_{C_0C_1}$
}\label{Fig_SK_T01_g1}
\end{figure}

The super-operator $\mathcal{T}$ can be taken as a mapping between the operators. The transition operator $T_{C_0C_1}$ and its dual $\tilde{T}_{C_0C_1}$ are only special form of the mapping. Let us consider two density matrices $\rho_0$ and $\sigma_0$, one could define the operator
\bea
\mathcal{D}_{C_0C_1}:= \mathcal{T}(\rho_0\otimes \sigma_0).
\eea
By the defintion of super-operator, we have
\bea
\mathcal{D}_{C_0C_1}=U^\dagger|k\rangle \langle l|U \rho_0\otimes |l\rangle \langle k|\sigma_0.
\eea
Or we can write it as
\bea
&&\mathcal{D}_{C_0C_1}= \langle j| U^\dagger|k\rangle \langle l|U \rho_0|i\rangle |j\rangle \langle i| \otimes  \langle k|\sigma_0|m\rangle |l\rangle\langle m|\nn \\
&&\phantom{D(\rho_0,\sigma_0)}=\langle j| U^\dagger \sigma_0|k\rangle \langle l|U \rho_0|i\rangle |j\rangle \langle i| \otimes |l\rangle\langle k|.
\eea
Assume $\rho_0=|\psi_0\rangle \langle \psi_0|$ and $\sigma_0=|\phi_0\rangle \langle \phi_0|$, we have
\bea
\mathcal{D}_{C_0C_1}=\langle j| U^\dagger |\phi_0\rangle \langle \phi_0|k\rangle \langle l|U |\psi_0\rangle\langle \psi_0|i\rangle |j\rangle \langle i| \otimes |l\rangle\langle k|.
\eea
 The operator $\mathcal{D}_{C_0C_1}$ can be represented by path integral as shown in Fig.\ref{Fig_SK_T01_g1}. 

We would have the following relation
\bea
tr_{C_0}\mathcal{D}_{C_0C_1}=U\rho_0 U^\dagger\sigma_0,\quad tr_{C_1}\mathcal{D}_{C_0C_1}=U^\dagger\sigma_0 U\rho_0.
\eea
Further, we have
\bea
tr_{C_0C_1}\mathcal{D}_{C_0C_1}=\Tr(\rho(t)\sigma_0).
\eea
$\Tr(\rho\sigma)$ is the Hilbert-Schmidt inner product for the density matrix $\rho$ and $\sigma$. Thus the trace of $D(\rho_0,\sigma_0)$ can be used to charaterize the distance between the density matrix $\rho(t)$ and $\sigma_0$. For two pure states $\rho_0=|\phi\rangle \langle \phi|$ and $\sigma_0=|\psi\rangle \langle \psi|$. The trace of $D(\rho_0,\sigma_0)$ is the fidelity between the evolution state $|\phi(t)\rangle=U|\phi\rangle$ and $|\psi\rangle$. Further, if $|\psi\rangle=|\phi\rangle$, the result is related to the return amptitude for the initial state $|\phi\rangle$.The operator $D(\rho_0,\sigma_0)$ is a useful quantities to study the thermalization process if one choose $\sigma_0$ to be the thermal state $\sigma_0=\frac{1}{Z}\sum_i e^{-\beta H}$. Its trace can be seen as a quantites to charaterize how close between the $\rho(t)$ and the thermal state. The trace of $D(\rho_0,\sigma_0)$ only include part information of this operator. We expect the eigenvalues of $D(\rho_0,\sigma_0)$ should include more information on the dynamics property, which may be more useful to study the thermalization progress in a quantum system.

In the studies of quantum quench, we usually focus on a subsystem, thus one could define a reduced operator for $D(\rho_0,\sigma_0)$, and then use the reduced operators to calculate the entropy-related quantities, which may be more useful to charaterize the quantum quench progress in a quantum system \cite{Calabrese:2016xau}.

In Section~\ref{Section_generalization} we introduced the generalization $\mathcal{F}_{C_0C_1,s}$ and its Hermitian conjugate $\mathcal{F}_{C_0C_1,s}^\dagger$. Let us define two density matrices $\rho_0=|\psi_0\rangle \langle \psi_0|$ and $\rho_1=|\psi_1\rangle \langle \psi_1|$. By comparing Fig.~\ref{Fig_SK_T01_g1} with Figs.~\ref{Fig_SK_T01_g3_HJ} and \ref{Fig_SK_T01_g3}, we find the following relation:
\bea
\mathcal{D}(\rho_0,\rho_1)=\tr_{C_1}(\mathcal{F}_{C_0C_1,s}) \otimes \tr_{C_0}(\mathcal{F}_{C_0C_1,s}^\dagger),
\eea
which can also be obtained by using the expressions of these operators.
\section{Moments of $T_{C_0C_1...C_i...C_{N-1}}$}\label{Section_moments_any_slices}
In this section we would like to show more examples of the moments of the transition operator $T_{C_0C_1...C_i}$. We will choose the initial density matrix to be the thermal state. The following two figures show the $tr(T_{C_0C_1}^3)$ and $tr(T_{C_0C_1}^4)$. One could draw higher-order moments for $T_{C_0C_1}$. We can see they satisfy the law (\ref{n_moment}).
\begin{figure}[htbp]
  \centering
  \includegraphics[width=0.76\textwidth]{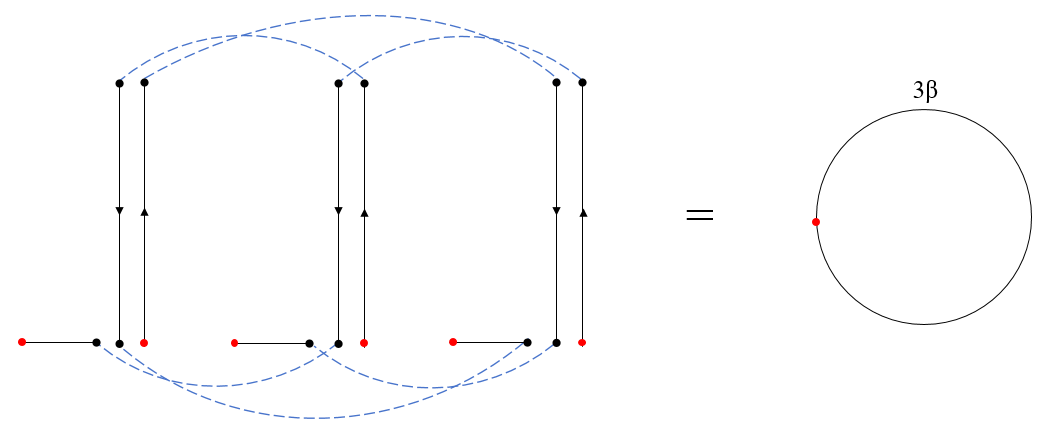}
  \caption{Path integral representation $trT_{C_0C_1}^3$. The circle on the right hand side of equation refer to the path integral on imaginary time direction, the circumference is $3\beta$.
}\label{Fig_trace_T01_3moment}
\end{figure}
\begin{figure}[htbp]
  \centering
  \includegraphics[width=0.76\textwidth]{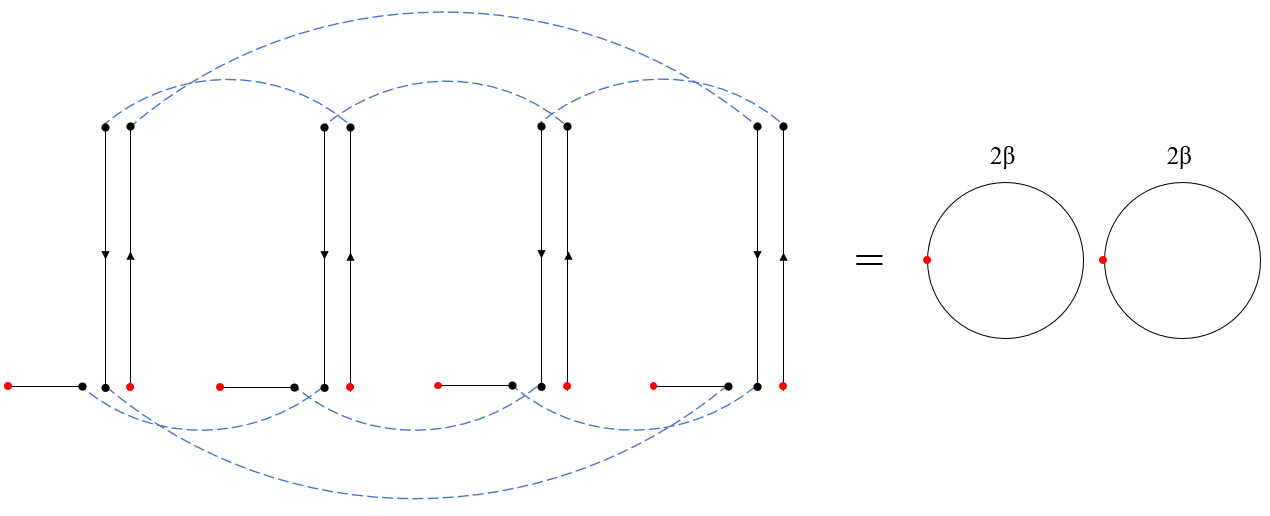}
  \caption{Path integral representation $trT_{C_0C_1}^4$.}\label{Fig_trace_T01_4moment}
\end{figure}

\begin{figure}[htbp]
  \centering
  \includegraphics[width=0.8\textwidth]{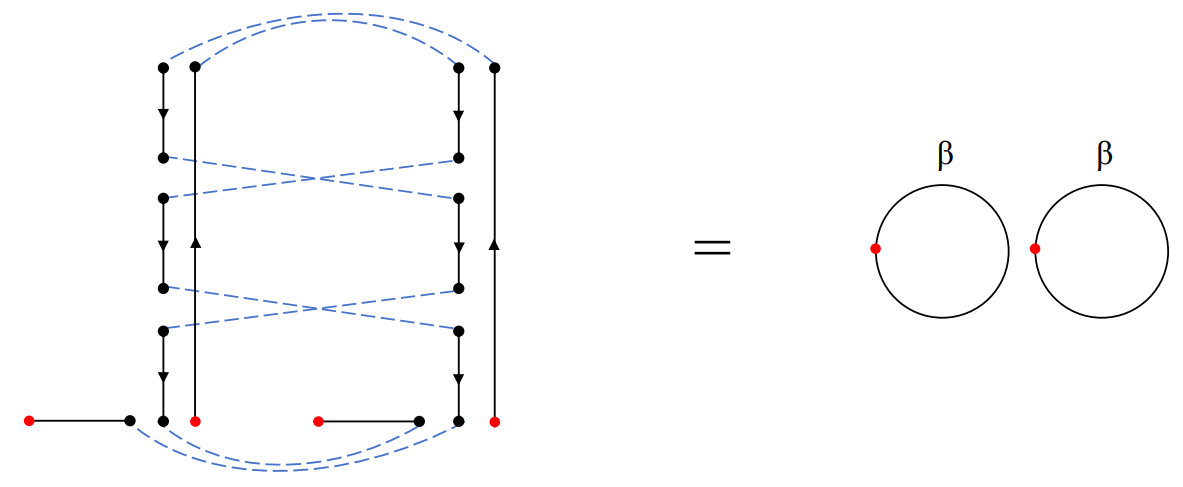}
  \caption{Path integral representation $trT_{C_0C_1C_2C_3}^2$. 
}\label{Fig_trace_T0123_2moment}
\end{figure}

\begin{figure}[htbp]
  \centering
  \includegraphics[width=0.8\textwidth]{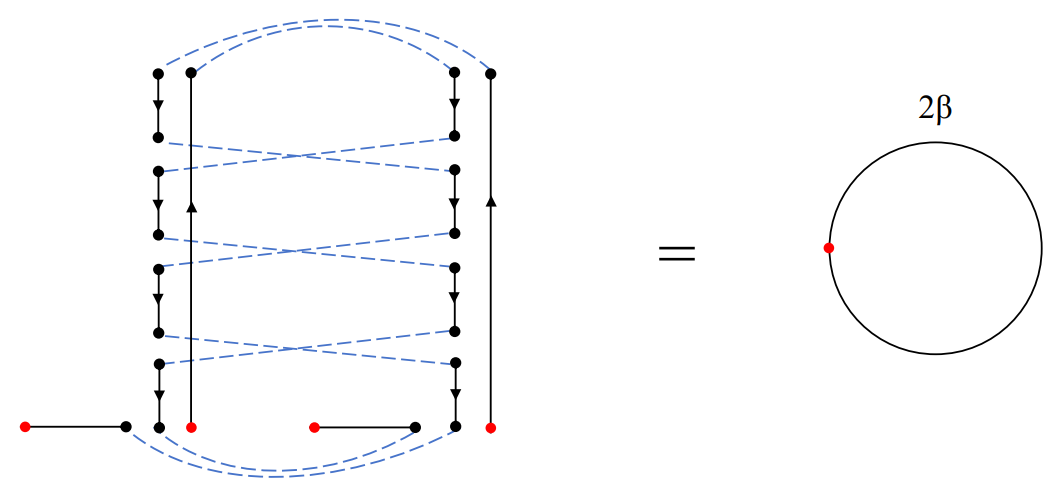}
  \caption{Path integral representation $tr T_{C_0C_1C_2C_3C_4}^2$. 
}\label{Fig_trace_T01234_2moment}
\end{figure}

It is also straightforward to compute the moment for $T_{C_0C_1...C_i...C_{N-1}}$. Here we only show some examples in Fig.\ref{Fig_trace_T0123_2moment} and Fig.\ref{Fig_trace_T01234_2moment}. The $2$-nd moment for $T_{C_0C_1...C_i...C_{N-1}}$ is summarized in Eq.(\ref{2moment_any_slices}).


\begin{figure}[htbp]
  \centering
  \includegraphics[width=0.7\textwidth]{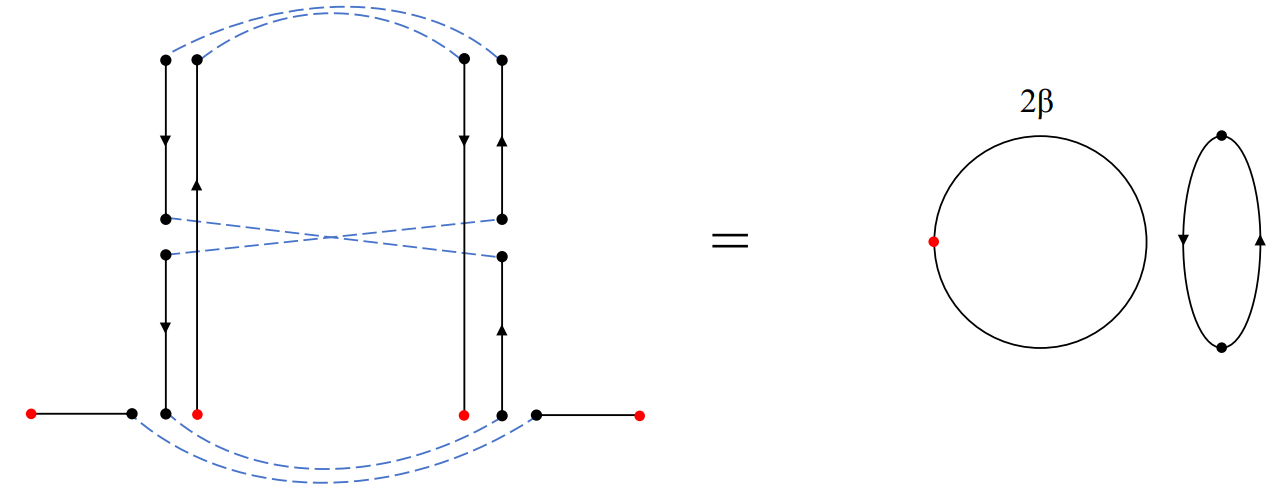}
  \caption{Path integral representation $trT_{C_0C_1C_2}^2$. 
}\label{Fig_trace_T012_2moment_TTHJ}
\end{figure}

\begin{figure}[htbp]
  \centering
  \includegraphics[width=0.7\textwidth]{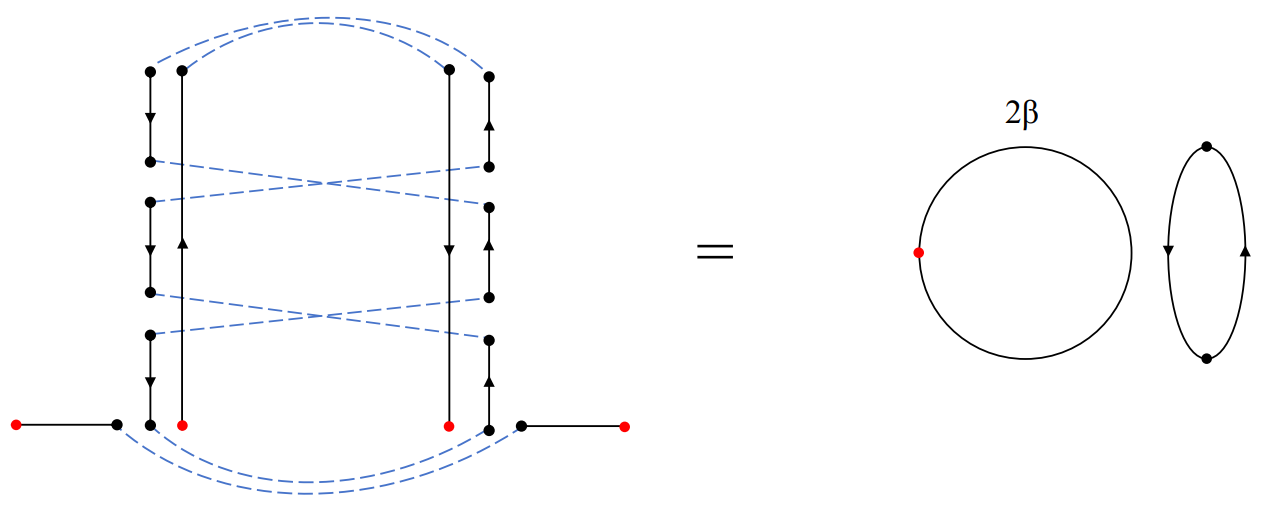}
  \caption{Path integral representation $tr(T_{C_0C_1C_2C_3}T_{C_0C_1C_2C_3}^\dagger)$. 
}\label{Fig_trace_T0123_2moment_TTHJ}
\end{figure}

\begin{figure}[htbp]
  \centering
  \includegraphics[width=0.7\textwidth]{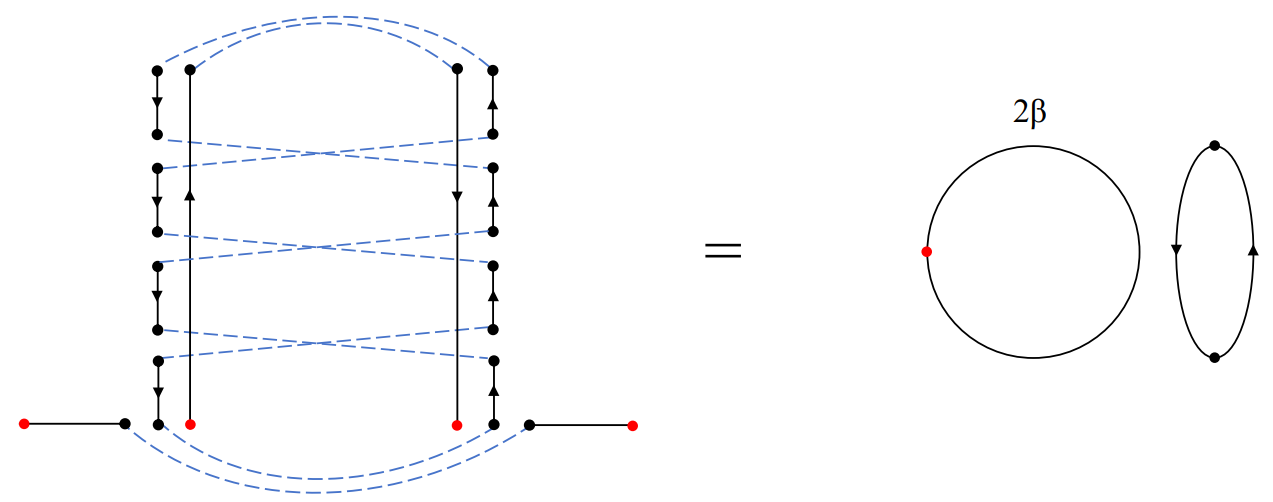}
  \caption{Path integral representation $tr( T_{C_0C_1C_2C_3C_4}T_{C_0C_1C_2C_3C_4}^\dagger)$. 
}\label{Fig_trace_T01234_2moment_TTHJ}
\end{figure}

In Section.\ref{Section_moment_transition operator} we use the digram to show how to compute $tr(T_{C_0C_1}T_{C_0C_1}^\dagger)$. We can also compute the higher-power. Fig.\ref{Fig_trace_T01_4moment} is an example. We also show the moments $tr(T_{C_0C_1C_2}T_{C_0C_1C_2}^\dagger)$, $tr(T_{C_0C_1C_2C_3}T_{C_0C_1C_2C_3}^\dagger)$ and $tr(T_{C_0C_1C_2C_3C_4}T_{C_0C_1C_2C_3C_4}^\dagger)$ in Fig.~\ref{Fig_trace_T012_2moment_TTHJ}, Fig.~\ref{Fig_trace_T0123_2moment_TTHJ} and Fig.~\ref{Fig_trace_T01234_2moment_TTHJ}. They all give the result $d~ \Tr \rho_0^2 $. The result can be generalized to any $T_{C_0C_1...C_i...C_{N-1}}$.

\begin{figure}[htbp]
  \centering
  \includegraphics[width=0.8\textwidth]{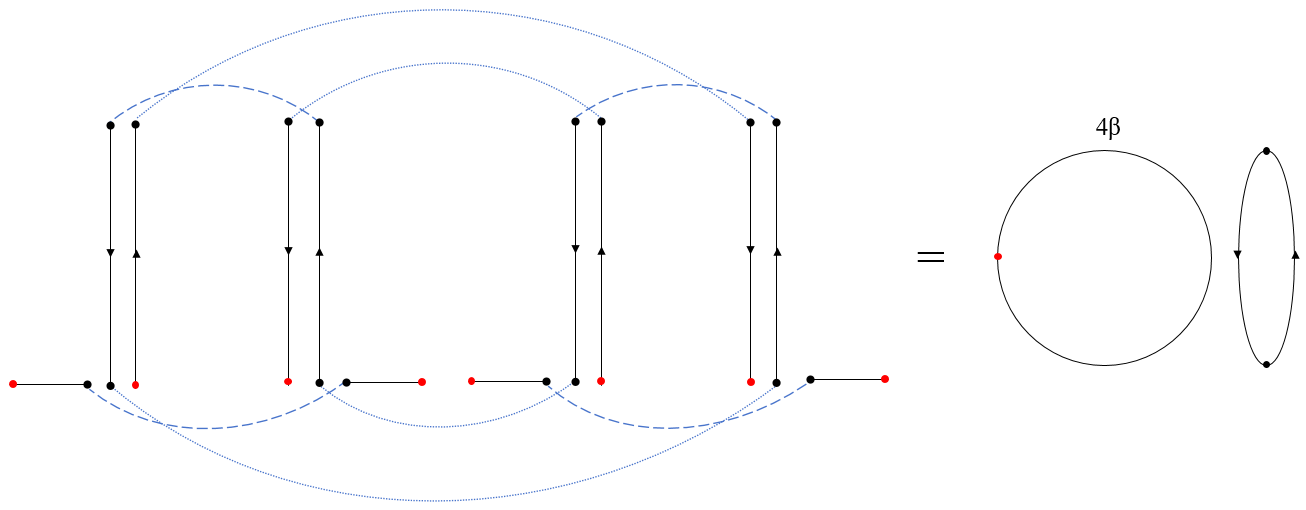}
  \caption{Path integral representation $tr(T_{C_0C_1}T_{C_0C_1}^\dagger)^2$.}\label{Fig_trace_T01_4moment_TTHJ}
\end{figure}


\section{Moments for reduced transition operator}\label{Appendix_moments_reduced}
In this section we will consider the moments:
\bea
tr (T_{A_0A_1}T_{A_0A_1}^\dagger),\quad tr (T_{A_0\bar A_1}T_{A_0\bar A_1}^\dagger).
\eea
We will only consider the term of order $O(\delta t)$. We have
\bea
tr (T_{A_0A_1}T_{A_0A_1}^\dagger)=tr (T^{(0)}_{A_0A_1}(T^{(0)}_{A_0A_1})^\dagger)+tr(T^{(0)}_{A_0A_1} (T^{(1)}_{A_0A_1})^\dagger)+tr(T^{(1)}_{A_0A_1} (T^{(0)}_{A_0A_1})^\dagger)+\cdots,
\eea
with
\bea
&&tr (T^{(0)}_{A_0A_1}(T^{(0)}_{A_0A_1})^\dagger)=\Tr\rho_{A_0}^2 d_{A_0},\nn\\
&&tr(T^{(0)}_{A_0A_1} (T^{(1)}_{A_0A_1})^\dagger)=-i\delta t \Tr_{\bar A_0} [\Tr_{A_0}(\rho_{A_0}\rho_0)\Tr_{A_0}H]\nn \\
&&\phantom{tr(T^{(0)}_{A_0A_1} (T^{(1)}_{A_0A_1})^\dagger)=}+i\delta t d_{A_0} \Tr_{A_0} [\rho_{A_0} \Tr_{\bar A_0}(\rho_0H)]\nn \\
&&tr(T^{(1)}_{A_0A_1} (T^{(0)}_{A_0A_1})^\dagger)=\left[tr(T^{(0)}_{A_0A_1} (T^{(1)}_{A_0A_1})^\dagger)\right]^*,
\eea
where $d_{A_0}$ is the dimension of the Hilbert space $\mathcal{H}_{A_0}$.

Simliarly, for $tr (T_{A_0\bar A_1}T_{A_0\bar A_1}^\dagger)$ we have
\bea
tr (T_{A_0\bar A_1}T_{A_0\bar A_1}^\dagger)=tr (T^{(0)}_{A_0\bar A_1}(T^{(0)}_{A_0\bar A_1})^\dagger)+tr(T^{(0)}_{A_0\bar A_1} (T^{(1)}_{A_0\bar A_1})^\dagger)+tr(T^{(1)}_{A_0\bar A_1} (T^{(0)}_{A_0\bar A_1})^\dagger)+\cdots,
\eea
with
\bea
&&tr (T^{(0)}_{A_0\bar A_1}(T^{(0)}_{A_0\bar A_1})^\dagger)=\Tr\rho_0^2,\nn \\
&&tr(T^{(0)}_{A_0\bar A_1} (T^{(1)}_{A_0\bar A_1})^\dagger)=-i\delta t \Tr_{A_0} \big[\Tr_{\bar A_0} (\rho_0|\bar L\rangle \langle \bar K|H)\Tr_{\bar A_0}(\rho_0 |\bar K\rangle \langle \bar L|) \big]\nn \\
&&\phantom{tr(T^{(0)}_{A_0\bar A_1} (T^{(1)}_{A_0\bar A_1})^\dagger)=}+i\delta t \Tr(\rho_0H)\nn \\
&&tr(T^{(1)}_{A_0\bar A_1} (T^{(0)}_{A_0\bar A_1})^\dagger)=\big[ tr(T^{(0)}_{A_0\bar A_1} (T^{(1)}_{A_0\bar A_1})^\dagger)\big]^*.
\eea
\subsection{Two qubits system}\label{Appendix_twoqubits}
In this section, we present several formulas for the two qubits system.. For the initial state $|\psi_0\rangle =\sin\theta |01\rangle +\cos\theta |10\rangle$, we have
\bea
T_{A_0A_1}=\left(
\begin{array}{cccc}
 t_{11} & 0 & 0 & 0 \\
 0 & t_{22} & t_{23}& 0 \\
 0 & t_{32} & t_{33} & 0 \\
 0 & 0 & 0 & t_{44} \\
\end{array}
\right),
\eea
with
\bea
&&t_{11}=\left(\frac{1}{2}+\frac{i}{2}\right) \cos (2 J t) \sin (\theta ) (i \sin (2 J t-\theta )+\sin (2 J t+\theta )),\nn \\
&&t_{22}=\left(-\frac{1}{2}-\frac{i}{2}\right) (i \cos (2 J t-\theta )+\cos (2 J t+\theta )) \sin (2 J t) \sin (\theta ),\nn \\
&&t_{23}=\frac{1}{2} \cos (\theta ) \left(\cos (\theta )+e^{-4 i J t} (\cos (\theta )-\sin (\theta ))+\sin (\theta )\right), \nn \\
&&t_{32}=\frac{1}{2} \sin (\theta ) \left(\cos (\theta )+\sin (\theta )+e^{-4 i J t} (\sin (\theta )-\cos (\theta ))\right),\nn \\
&&t_{33}=\left(\frac{1}{2}+\frac{i}{2}\right) \cos (\theta ) \sin (2 J t) (\sin (2 J t-\theta )-i \sin (2 J t+\theta )),\nn\\
&&t_{44}=\left(\frac{1}{2}+\frac{i}{2}\right) \cos (2 J t) \cos (\theta ) (\cos (2 J t-\theta )-i \cos (2 J t+\theta )),
\eea
and
\bea
T_{A_0B_1}=\left(
\begin{array}{cccc}
 t_{11} & 0 & 0 & 0 \\
 0 & t_{22} & t_{23} & 0 \\
 0 & t_{32} & t_{33} & 0 \\
 0 & 0 & 0 & t_{44} \\
\end{array}
\right)
\eea
with
\bea
&&t_{11}=\left(-\frac{1}{2}-\frac{i}{2}\right) (i \cos (2 J t-\theta )+\cos (2 J t+\theta )) \sin (2 J t) \sin (\theta ),\nn \\
&&t_{22}=\left(\frac{1}{2}+\frac{i}{2}\right) \cos (2 J t) \sin (\theta ) (i \sin (2 J t-\theta )+\sin (2 J t+\theta )),\nn \\
&&t_{23}=\frac{1}{2} \cos (\theta ) \left(\cos (\theta )+\sin (\theta )+e^{-4 i J t} (\sin (\theta )-\cos (\theta ))\right),\nn \\
&&t_{32}=\frac{1}{2} \sin (\theta ) \left(\cos (\theta )+e^{-4 i J t} (\cos (\theta )-\sin (\theta ))+\sin (\theta )\right),\nn \\
&&t_{33}=\left(\frac{1}{2}+\frac{i}{2}\right) \cos (2 J t) \cos (\theta ) (\cos (2 J t-\theta )-i \cos (2 J t+\theta )),\nn \\
&&t_{44}=\left(\frac{1}{2}+\frac{i}{2}\right) \cos (\theta ) \sin (2 J t) (\sin (2 J t-\theta )-i \sin (2 J t+\theta )).
\eea
In general, they are non-Hermitian matrices.

\section{Moments for thermal field double state}\label{Appendix_TFD}
In this section we would like to show more examples for the moments of $\mathcal{F}_{L_0L_1,s}^\dagger$ and $\mathcal{F}_{L_0R_1,s}^\dagger$.
\begin{figure}[htbp]
  \centering
  \includegraphics[width=0.8\textwidth]{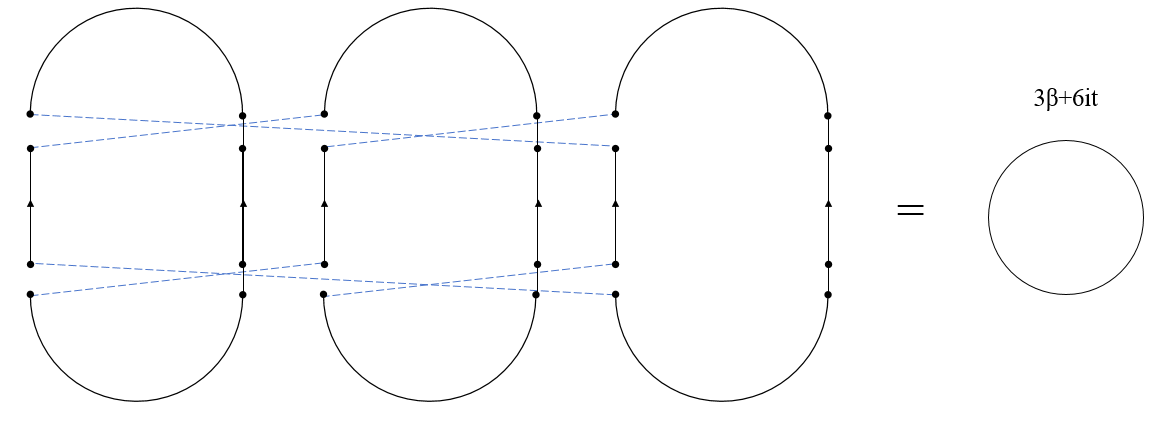}
  \caption{Path integral representation $tr( \mathcal{F}_{L_0L_1,s}^\dagger)^3$. 
}\label{Fig_TFD_LL_3moment}
\end{figure}
\begin{figure}[htbp]
  \centering
  \includegraphics[width=0.8\textwidth]{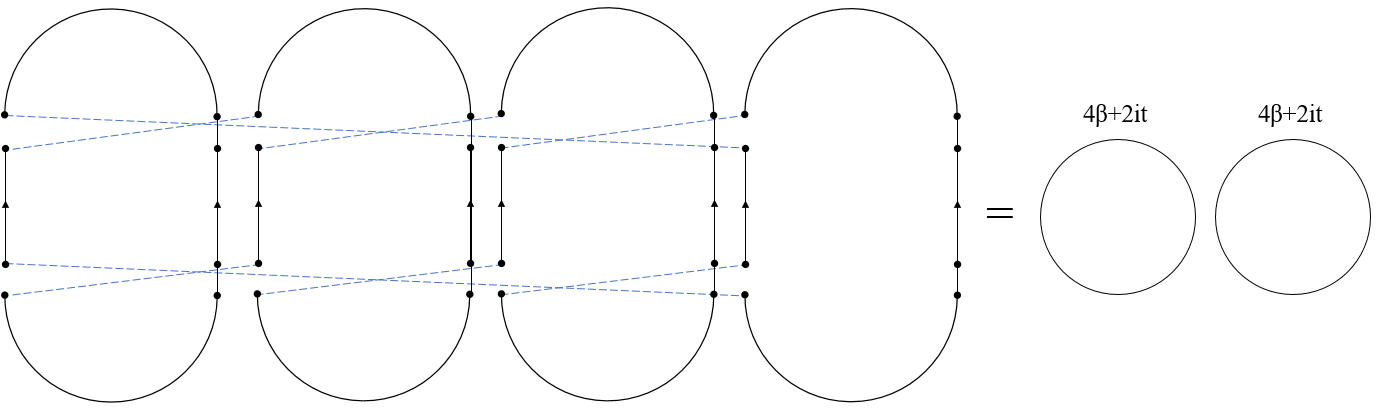}
  \caption{Path integral representation $tr( \mathcal{F}_{L_0L_1,s}^\dagger)^4$. 
}\label{Fig_TFD_LL_4moment}
\end{figure}
\begin{figure}[htbp]
  \centering
  \includegraphics[width=0.8\textwidth]{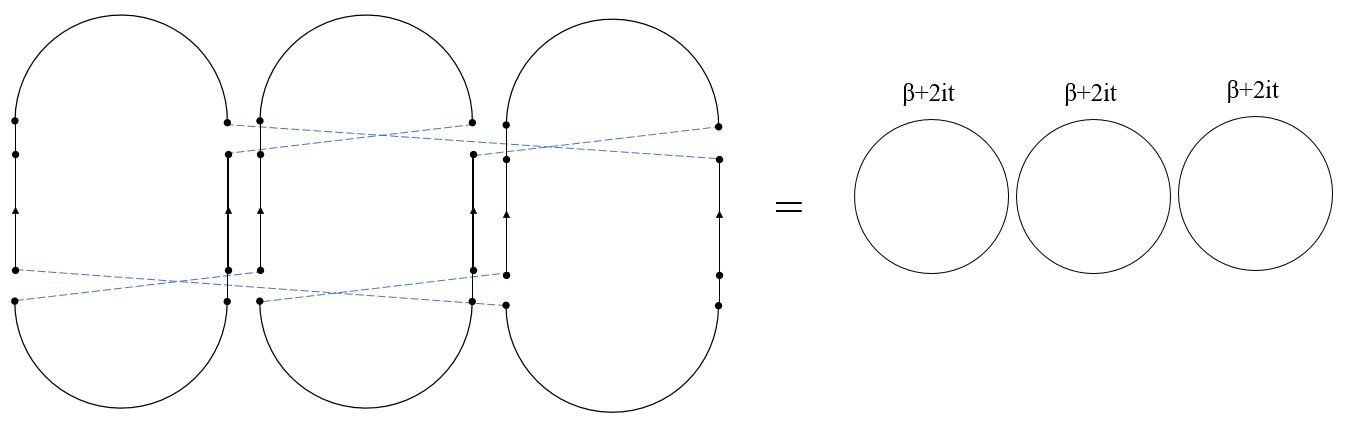}
  \caption{Path integral representation $tr( \mathcal{F}_{L_0R_1,s}^\dagger)^3$. 
}\label{Fig_TFD_LR_3moment}
\end{figure}

\end{document}